\documentclass[10pt,conference,compsoc]{IEEEtran}
\usepackage{ifpdf}
\usepackage{cite}
\ifCLASSINFOpdf
  \usepackage[pdftex]{graphicx}
  \graphicspath{{./figs/}{./plots/}}
  \DeclareGraphicsExtensions{.pdf,.jpeg,.png}
  \pdfpageattr{/Group <</S /Transparency /I true /CS /DeviceRGB>>}
\else
  \usepackage[dvips]{graphicx}
  \graphicspath{{./figs/}{./plots/}}
  \DeclareGraphicsExtensions{.eps}
\fi
\usepackage{amsmath}
\ifCLASSOPTIONcompsoc
  \usepackage[caption=false,font=normalsize,labelfont=sf,textfont=sf]{subfig}
\else
  \usepackage[caption=false,font=footnotesize]{subfig}
\fi
\usepackage{url}
\usepackage{tikz}
\usetikzlibrary{patterns, shapes, positioning, decorations.pathreplacing,
backgrounds, fit}
\usepackage{gnuplot-lua-tikz}
\usepackage{forest}
\usepackage{tcolorbox}
\usepackage{textcomp}
\usepackage[binary-units]{siunitx}
\providecommand{\mbps}[1]{\SI{#1}{\mega\byte\per\second}}

\providecommand{\s}[1]{\SI{#1}{\second}}

\begin{document}
\title{DAOS for Extreme-scale Systems in Scientific Applications}

%\author{\IEEEauthorblockN{M. Scot Breitenfeld, Neil Fortner, Jordan Henderson, Jerome Soumagne}
%\IEEEauthorblockA{The HDF Group\\
%Champaign, IL 61820}
%\and
%\IEEEauthorblockN{Quincey Koziol}
%\IEEEauthorblockA{Lawrence Berkeley National Laboratory\\
%Berkeley, CA 94720}
%\and
%\IEEEauthorblockN{Mohamad Chaarawi}
%\IEEEauthorblockA{Intel\\
%Address
%}
%}


\author{\IEEEauthorblockN{M. Scot Breitenfeld\IEEEauthorrefmark{1},
Neil Fortner\IEEEauthorrefmark{1},
Jordan Henderson\IEEEauthorrefmark{1}, 
Jerome Soumagne\IEEEauthorrefmark{1},\\
Mohamad Chaarawi\IEEEauthorrefmark{2}, Johann Lombardi\IEEEauthorrefmark{2} and Quincey Koziol\IEEEauthorrefmark{3}}
\IEEEauthorblockA{\IEEEauthorrefmark{1}The HDF Group,
Champaign, IL 61820}
\IEEEauthorblockA{\IEEEauthorrefmark{2}Intel Corporation,
Santa Clara, CA 95054}
\IEEEauthorblockA{\IEEEauthorrefmark{3}Lawrence Berkeley National Laboratory,
Berkeley, CA 94720}
}


\maketitle

\begin{abstract}
Exascale I/O initiatives will require new and fully integrated I/O models which are capable of providing straightforward functionality, fault tolerance and efficiency. One solution is the Distributed Asynchronous Object Storage (DAOS) technology, which is primarily designed to handle the next generation NVRAM and NVMe technologies envisioned for providing a high bandwidth/IOPS storage tier close to the compute nodes in an HPC system. In conjunction with DAOS, the HDF5 library, an  I/O library for scientific applications, will support end-to-end data integrity, fault tolerance, object mapping, index building and querying. This paper details the implementation and performance of the HDF5 library built over DAOS by using three representative scientific application codes.
\end{abstract}

\begin{IEEEkeywords}
I/O Software Stack, Storage, Resilience, Exascale, Parallel File System,
High Performance Computing, DAOS, HDF5 
\end{IEEEkeywords}

% \fi
\IEEEpeerreviewmaketitle

\section{Introduction}

Scientists, engineers and application developers may soon need to
address the I/O challenges of computing on future exaflop machines.
Economic realities drive the architecture, performance, and reliability
of the hardware that will comprise an exascale I/O system~\cite{Kogge2008}.
Moreover, I/O researchers~\cite{Isaila2016} have highlighted
significant weaknesses in the current I/O stacks that will need to
be addressed in order to enable the development of systems that measurably
demonstrate all of the properties required of an exascale I/O system.
Possible poor filesystem performance at exascale has lead to
the introduction of new or augmented file systems~\cite{Mehta2012}.
It is also anticipated that failure will be the norm~\cite{Kogge2008}
and the I/O system as a whole will have to handle it as transparently
as possible, all while providing efficient, sustained, scalable and
predictable I/O performance. The enormous quantities of data, and
especially of application metadata, envisaged at exascale will become
intractable if there can be no assurance of consistency in the face
of all possible recoverable failures and if there can be no assurance
of error detection in the face of all possible failures.

Furthermore, HPC applications developers and scientists need to be
able to think about their simulation models at higher levels of abstraction
if they are to be free to work effectively on problems of the size and complexity
that become possible at exascale. This, in turn, puts
pressure on I/O APIs to become more expressive by describing high-level
data objects, their properties and relationships. Additionally, HPC
developers and scientists must be able to interact with, explore and
debug their simulation models. The I/O APIs should, therefore, support
index building and traversal, and be integrated with
a high level interpreted language such as Python to permit ad-hoc
programmed queries. Currently, high-level HPC I/O libraries support
relatively static data models and provide little or no support for
efficient ad-hoc querying and analysis. 

To provide
a high degree of flexibility and portability to the user, the HDF5
library~\cite{Folk2011}\cite{Folk1999} organizes data into a hierarchical tree
that is composed of groups, datasets and attributes. Groups and datasets are linked
and defined by a link name; attributes are attached to these objects and defined
by a name and value. Datasets store the actual data and may be
contiguously  mapped from an application memory to a file, or stored in more
complex patterns to ease further access and analysis of the data.
This paper discusses an effort to port applications using HDF5 to use an exascale transactional I/O stack. Section \ref{sec:Exascale-I/O-challenges}
gives an overview of the envisioned exascale I/O systems and the challenges
associated with them. Section \ref{sec:A-new-I/O} discusses a new
transactional storage I/O stack implementation via HDF5, and Section
\ref{sec:Application-I/O-strategies} gives the strategies involved
in porting scientific applications to the proposed storage I/O stack
and gives numerous benchmarks for each application.

\section{Exascale I/O Challenges\label{sec:Exascale-I/O-challenges}}

One possible approach for application I/O at exascale is to become object
oriented. Meaning, rather than reading and writing files, applications
will instantiate and persist rich distributed data structures using
a transactional mechanism~\cite{Lofstead2013}. As concurrency increases
by orders of magnitude, programming styles will be forced to become
more asynchronous~\cite{Keyes2011} and I/O APIs will have to take
a lesson from HPC communications libraries, by using non-blocking
operations to initiate I/O.
I/O subsystems that impose unnecessary serialization on applications
(e.g., by providing over-ambitious guarantees on the resolution of conflicting
operations) simply may not scale. It could, therefore, become the
responsibility of the I/O system to provide, rather than impose, appropriate
scalable mechanisms to resolve such conflicts. It is then the responsibility
of the application to use those mechanisms correctly.

Components and subsystems in the numbers that will be deployed at
exascale mean that failures are unavoidable and relatively frequent.
Recovery must be designed into the I/O stack from the ground up and
applications must be provided with APIs that enable them to recover
cleanly and quickly when failures cannot be handled transparently.
This mandates a transactional I/O model such that applications can be
guaranteed their persistent data models remain consistent in the face
of all possible failures. Recovery should also guarantee consistency for
redundant object data and filesystem metadata whether such mechanisms
are implemented within the filesystem or in middleware. Such behavior can be achieved by confining
the object namespace within containers that appear in the filesystem
namespace as single files. Higher levels of the I/O stack will see
these containers as private scalable object stores, driving the need
for a new standard low-level I/O API to replace POSIX for these containers.
This provides a common foundation for alternative middleware stacks
and high-level I/O models, suitable to different application domains.

\subsection{Exascale System I/O Architecture}

The economics and performance tradeoffs between disk and solid state
persistent storage or NVRAM determine much of the exascale system
architecture. NVRAM is required to address performance issues but
cannot scale economically to the volumes of data anticipated. Conversely,
disks can address the volume of data but not the economical aspects of the performance
requirements.

\begin{figure}
\begin{centering}
\includegraphics[width=1\linewidth]{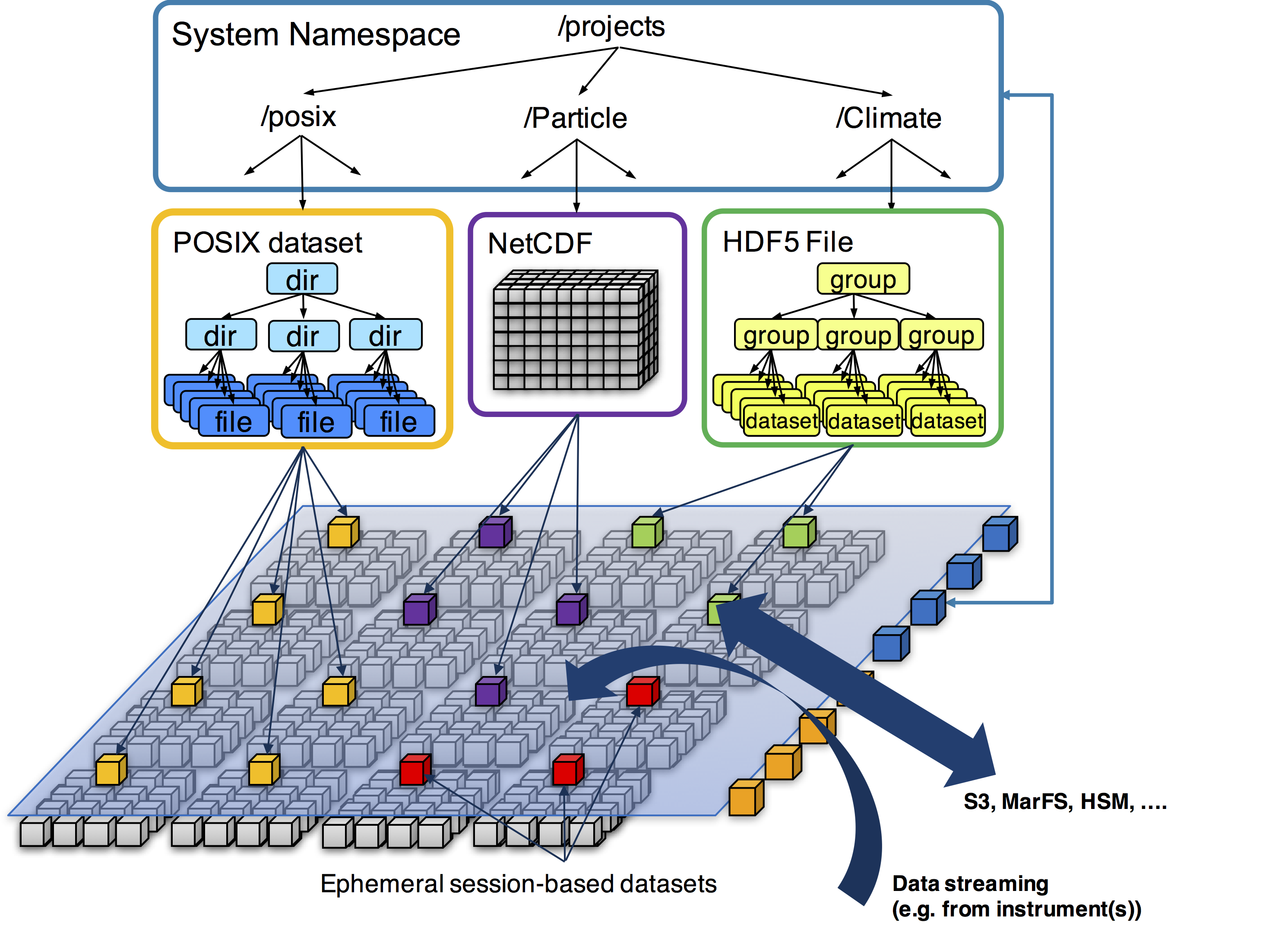} 
\par\end{centering}
\vspace{-5pt}
\caption{\label{fig:Exascale-I/O-architecture}Vision for exascale storage.}
\end{figure}

Economics dictates a HPC cluster, Fig. \ref{fig:Exascale-I/O-architecture}, with hundreds of thousands of compute nodes interconnected with a scalable, high-speed, low-latency fabric where all (or a subset) of the nodes, called storage nodes, have direct access to byte-addressable persistent memory and optionally block-based NVMe storage as well. A storage node can export over the network one or more object, each of which corresponds to a fixed-size partition of its directly accessible storage. The goal of the HPC cluster is to have both fault tolerance and concurrency mechanisms. A storage node can host multiple objects within the limits of the available storage capacity.

\subsection{Exascale Compute Cluster}

Typically, I/O nodes (ION) will run Linux and have direct access to the global
shared filesystem. Each ION will serve a different set of compute
nodes (CN) to ensure I/O communications between CNs and IONs exit the exascale
network as fast as possible. The NVRAM on the IONs will provide a
key-value store for use as a pre-staging cache and a hot storage tier  to handle peak
I/O load and defensive I/O. Write data captured by the hot storage tier, will be
repackaged by a layout optimizer according to expected usage into
objects sized to match the bandwidth and latency properties of the
storage tier targeted on the shared global filesystem. These storage
objects will then be written in redundant groups using erasure codes
or mirroring as appropriate. Object placement will be dynamic and
responsive to server load to ensure servers remain evenly balanced
and throughput is maximized.

CN or ION failure will be handled transparently by restarting the
application from the last accessible checkpoint. In the case of CN
failure, or if the NVRAM subsystem used for the hot storage tier is highly
available and reliable (i.e., fully redundant and accessible via multiple
paths) this will only require rollback to the last checkpoint stored
in the hot storage tier. Otherwise the application will have to restart
from the last checkpoint saved to the global shared filesystem.

\section{A New I/O Software Stack\label{sec:A-new-I/O}}

\begin{figure}
\begin{centering}
\includegraphics[width=1\linewidth]{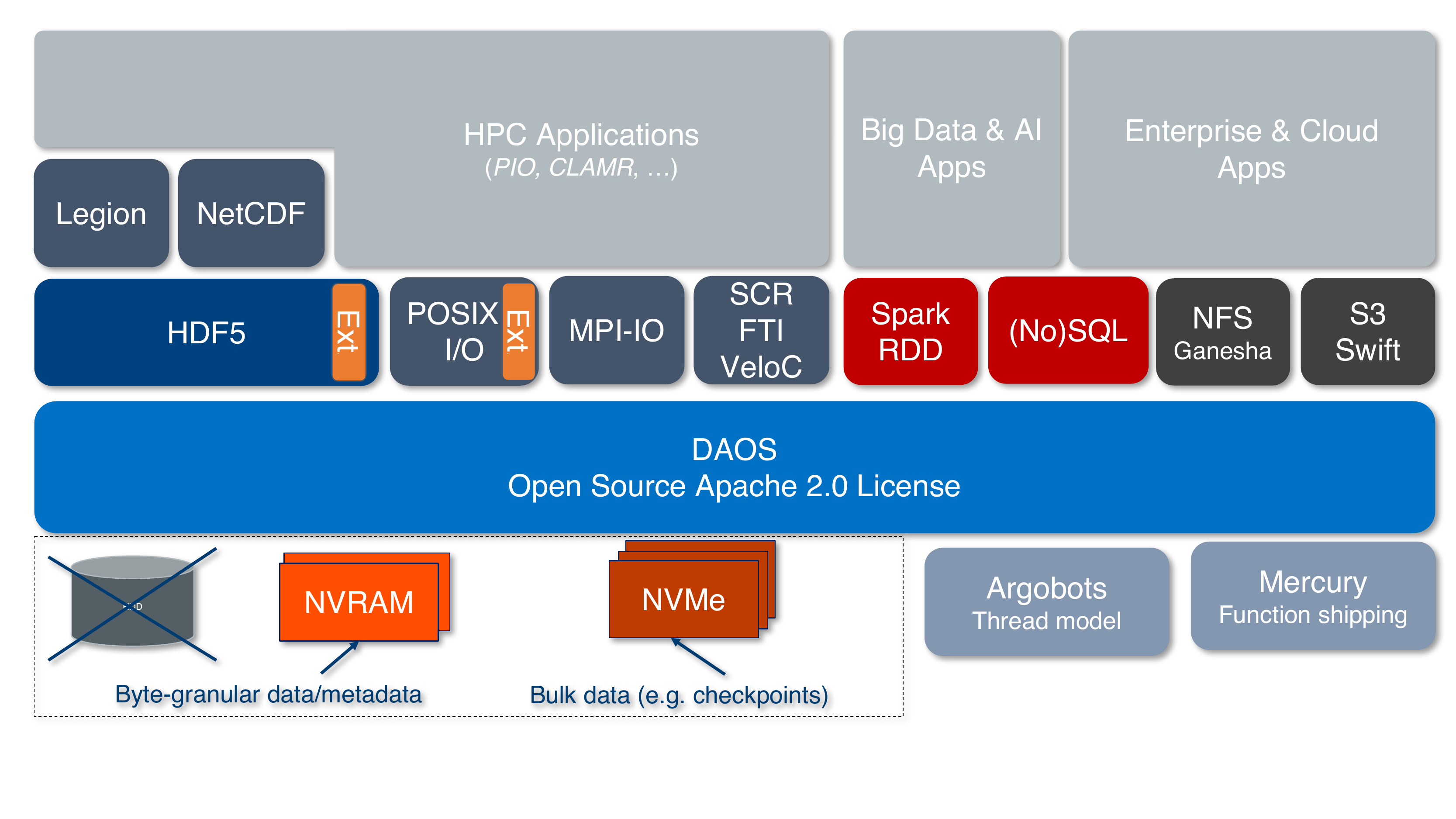} 
\par\end{centering}
\vspace{-15pt}
\caption{\label{fig:FF-stack-configuration}DAOS stack configuration.}
\end{figure}

New HDF5 object storage APIs were developed to add
support for end-to-end data integrity, fault tolerance, object mapping,
index building and query. The
new I/O APIs are implemented in the HDF5 library and provide a layer
over the lower level object storage APIs. The top of the stack features a new version of HDF5, which directly
interfaces with DAOS, Distributed Asynchronous Object Storage, and
provides scalable, transactional object storage containers for encapsulating
entire exascale datasets and their metadata, Fig. \ref{fig:FF-stack-configuration}.

\begin{figure}
\begin{centering}
\includegraphics[width=1\linewidth]{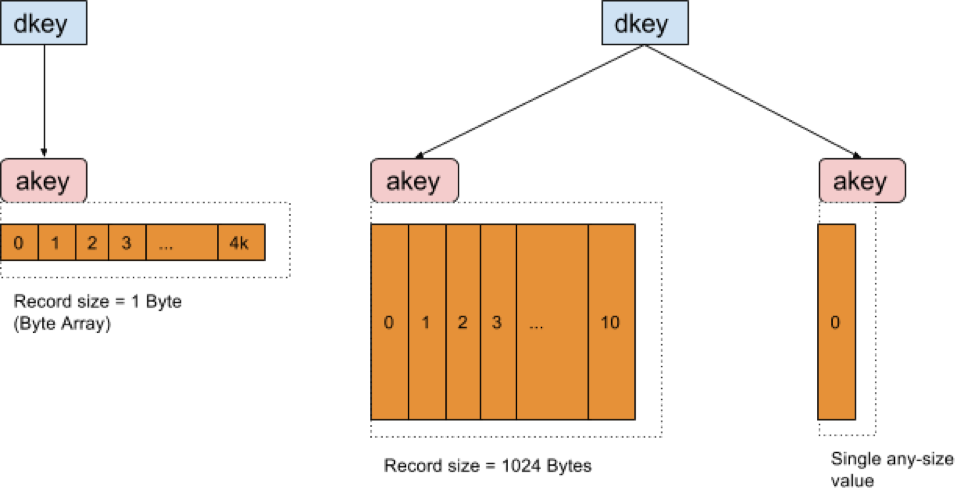} 
\par\end{centering}
\vspace{-5pt}
\caption{\label{fig:DaosObj}DAOS object model.}
\end{figure}

The DAOS storage system itself builds on existing techniques~\cite{Carns2016}~\cite{Carns2015}
and middleware such as Argobots~\cite{Seo2016} for fast user-level threading
and Mercury~\cite{Soumagne2013} for low-latency messaging and high-bandwidth data
transfers.
The essence of the DAOS storage model is a key-array object providing
efficient storage for both structured (fixed-size array element addressed
by index) and unstructured (variable length data stored in first array
index) data, Fig. \ref{fig:DaosObj}. The \textit{object} key is a 2-level key:
\begin{enumerate}
\item \textit{Distribution Key} (dkey) determines the placement. A
user groups data under a single dkey to hint locality and colocation
of that data on a single target;
\item \textit{Attribute Key} (akey) identifies an array of values.
\end{enumerate}
An array value is an arbitrary blob with an arbitrary size (from 1-byte
to many GBs).

Many object schemas (replication/erasure code, static/dynamic striping and others) are provided to achieve high availability and scalability. The schema framework is flexible and easily expandable to allow for new custom schema types in the future. The actual object layout is generated on open. The object class is extracted from the object identifier to determine the schema of the object to be opened. End-to-end integrity is assured by protecting both object data and metadata with checksums during network transfer and storage.

\subsection{\label{subsec:I/O-Transaction-model}I/O Transaction Model}

The primary goal of the DAOS transaction model is to guarantee data model consistency with highly concurrent workloads. Applications should be able to safely update the dataset in-place and rollback to a known consistent state on failure.

DAOS also introduces the concept of a container which represents an object address space inside a pool. A container is the basic unit of atomicity and versioning. Any time a container is opened, a handle for that container is returned to the user. All object operations are explicitly tagged by the caller with both the container handle and a transaction identifier called an epoch. Operations submitted against the same epoch and container handle are applied atomically to a container on a successful commit.

The DAOS transactional model allows applications to concurrently update a DAOS container through different transactional contexts by utilizing different container handles. All operations submitted with the same epoch (transaction number) and container handle are guaranteed to be atomically committed or aborted. Several applications or processes/threads within an application may independently open and access a container through different handles. DAOS tracks all I/O submitted with both the epoch state and the container handle.

This “all or nothing” semantic eliminates the possibility of partially integrated updates on a container handle. On a successful commit, an epoch is guaranteed to be immutable, durable and consistent. Unused committed and aborted epochs for a container may be aggregated to reclaim space utilized by overlapping writes and reduce metadata complexity. The user is responsible though for conflicts (for example updating overlapping extents of an array in the same epoch), as DAOS will not have information at aggregation time to resolve such conflicts.

DAOS uses the concept of \textit{transactions}, where one or more
processes in the calling program can participate in a transaction,
and there may be multiple transactions in progress in a container
at any given time. Transactions are numbered, and the calling program
is responsible for assigning transaction numbers in the DAOS stack.
Updates in the form of additions, deletions and modifications are
added to a transaction and not made directly to a container. Once
a transaction is committed, the updates in the transaction are applied
atomically to the container.

The basic HDF5 sequence of transaction operations on a container for opening and writing is: 
\begin{enumerate}
\item \textit{start} transaction \textit{N};
\item \textit{update} the container;
\item \textit{finish} transaction \textit{N}. 
\end{enumerate}
Transactions can be finished in any order, but they are committed
in strict numerical sequence. The application controls when a transaction
is committed through its assignment of transaction numbers in ``create
transaction / start transaction'' calls and the order in which transactions
are finished, aborted, or explicitly skipped.

The \textbf{version} of the container after transaction \emph{N} has
been committed is \emph{N}. An application reading this version of
the container will see the results from all committed transactions
up through and including \emph{N}.

The application can \textbf{persist} a container version, \emph{N},
causing the data (and metadata) for the container contents that are
in hot storage to be copied to DAOS and atomically committed to persistent
storage.

The application can request a \textbf{snapshot} of a container version
that has been persisted to DAOS. This makes a permanent entry in the
namespace (using a name supplied by the application) that can be used
to access that version of the container. The snapshot is independent
of further changes to the original container and behaves like any
other container from this point forward. It can be opened for write
and updated via the transaction mechanism (without affecting the contents
of the original container), it can be read, and it can be deleted.

\subsection{\label{subsec:HDF5-EFF-implementation}HDF5 DAOS Implementation}

HDF5 provides a set of user-level object abstractions
for organizing, saving, and accessing application data in a storage
container, such as groups for creating a hierarchy of objects and
datasets for storing multi-dimensional data. The HDF5 binary file
format is no longer used in DAOS. Instead, each HDF5 
object is now represented as a set of Key-value objects used to store
HDF5 metadata, replacing binary trees that index byte streams. A 
version of the HDF5 library that supports a Virtual Object Layer (VOL)
was used for DAOS. For this work, a specialized HDF5 DAOS
VOL plug-in interfaces to DAOS replaced the traditional HDF5 storage-to-byte-stream
binary format with storage-to-DAOS objects.

Caching and prefetching is handled by DAOS, rather than by the HDF5
library, with the HDF5/DAOS VOL server translating an application's
directives for HDF5 objects into directives for DAOS. Whereas HDF5
traditionally provided knobs for controlling cache size and policy,
and then tried to ``do the right thing'' with respect to maintaining
cached data, DAOS relies on explicit user directives, with the expectation
that written data may be analyzed by another job before being evicted
from the hot storage tier. In addition to the changes ``beneath'' the
existing HDF5 API, the DAOS HDF5 version supports features seen as
critical to future exascale storage needs: asynchronous operations,
end-to-end integrity checking, and data movement operations that enable
I/O to a hot storage tier. Additionally, DAOS HDF5 handles DAOS
transactions logistics internally (i.e., the schema discussed in Section
\ref{subsec:I/O-Transaction-model}), resulting instantly in the capability
to improve fault tolerance of data storage and allow near real-time
analysis for producer/consumer workloads. Finally, the DAOS HDF5 version
has exascale capabilities that are targeted to both current and future
users that include both query/view/index APIs to enable and accelerate data analysis and
a map object that augments the group and dataset objects.

\section{\label{sec:Application-I/O-strategies}Application I/O Strategies}

This section discusses the strategies involved in porting scientific
applications to DAOS. Section \ref{subsec:CLAMR-application} discusses
porting the application \emph{CLAMR} to use HDF5 DAOS and gives an
overview of the usability and capabilities from the perspective of
a typical application code. The second application \emph{Legion, }Section
\emph{\ref{subsec:Legion-parallel-programming},} demonstrates using
DAOS in a data-centric programming model. The last two applications
NetCDF-4 and PIO, Section \ref{subsec:HL}, are higher-level I/O libraries
built on top of HDF5 and show the utilization of a higher level of
abstraction to simplify an application's interaction with HDF5 and
DAOS, which in turn should ease the transition to DAOS.

A small Intel DAOS prototype cluster, \textsl{Boro}, was used for
all development and benchmarks described in Sections \ref{subsec:CLAMR-application}-\ref{subsec:HL}.
Boro consists of Intel Xeon Processor E5-2699 v3 with two CPUs per
node. The DRAM (Kingston KVR21R15S4/8) is being used in place of the
NVRAM (Fig. \ref{fig:Exascale-I/O-architecture}) and the final storage
stage uses Seagate Constellation ES.3 ST1000NM0033 HDD. Boro uses InfiniBand as its network backbone.

Before presenting the applications that were evaluated and ported to utilize the new I/O software stack, it is worth mentioning that both the DAOS library and the HDF5 DAOS backend are still in a prototyping phase. The primary goal of this research was to prove that utilizing the new I/O stack is doable and easy once a middleware library is properly designed on top of DAOS. Tuning for optimal performance in both HDF5 and DAOS was not done as part of this work.  Therefore it is unknown how the following issues would effect the performance.
\begin{itemize}
\item All the applications, except CLAMR, utilizing DAOS did not use HDF5 chunking layout for dataset storage. Thus, only a single DAOS server with one service thread for I/O is utilized because in this case an HDF5 dataset is mapped to one DAOS object distribution. This undoubtedly will result in a bottleneck once the number of I/Os are increased.
\item Communication to the DAOS server uses libfabric with TCP over InfiniBand, whereas communication to the Lustre server is done over InfiniBand verbs.
\item Lustre servers use spinning disks for storage, but DAOS used a \textit{tmpfs} file system over DRAM (since NVRAM is not available for this cluster). 
\item The MPI-I/O driver that HDF5 uses on Lustre will aggregate small I/Os at the client side before submitting I/O to the Lustre servers. No such aggregation is available yet in the HDF5 DAOS plugin nor at the DAOS client, which is something that will be explored in the future.
\end{itemize}

\subsection{CLAMR Application\label{subsec:CLAMR-application}}

\begin{figure}
\begin{centering}
\resizebox{\linewidth}{!}{\definecolor{fblue}{RGB}{92,144,192}
\definecolor{fgreen}{RGB}{34,162,70}

\definecolor{folderbg}{RGB}{124,166,198}
\definecolor{folderborder}{RGB}{110,144,169}

\def\Size{4pt}
\tikzset{
  folder/.pic={
    \filldraw[draw=folderborder,top color=folderbg!50,bottom color=folderbg]
      (-1.05*\Size,0.2\Size+5pt) rectangle ++(.75*\Size,-0.2\Size-5pt);  
    \filldraw[draw=folderborder,top color=folderbg!50,bottom color=folderbg]
      (-1.15*\Size,-\Size) rectangle (1.15*\Size,\Size);
  }
}

\begin{forest}
 where level=1{
    child anchor=west,
    !u.parent anchor=south,
    s sep=0pt,
    before computing xy={
      l*=.5,
    },
    edge path'={(!u.parent anchor) -- ++(0,-5pt) -| (.child anchor)},
    for tree={
      grow'=0
    }
  }{},
  where level=2{
    child anchor=west,
    parent anchor=south,
    anchor=west,
    calign=first,
    inner xsep=6pt,
    s sep=2pt,
    edge path={
      \noexpand\path [draw, \forestoption{edge}]
      (!u.south west) +(12.5pt,0) |- (.child anchor) pic {folder} \forestoption{edge label};
    },
    before typesetting nodes={
      if n=1
        {insert before={[,phantom]}}
        {}
    },
    fit=rectangle,
    before computing xy={l=25pt},
  }{},
  where level=3{
    child anchor=west,
    parent anchor=south,
    anchor=west,
    edge path={
      \noexpand\path [draw, \forestoption{edge}]
      (!u.south west) |- (.child anchor) \forestoption{edge label};
    },
    before typesetting nodes={
      if n=1
        {insert before={[,phantom]}}
        {}
    },
    before computing xy={l=10pt},
  }{},
  for tree={
    font=\ttfamily\scriptsize,
    edge={thick}
  }
[/
[
    [/bootstrap
      [double\_vals]
      [int\_vals]
    ]
    [/mesh,
      [double\_vals]
      [cpu\_timers]
      [gpu\_timers]
      [int\_dist\_vals]
      [int\_vals]
    ]
]
[
    [/default,
      [storage]
      [i]
      [j]
      [level]
      [H]
      [U]
      [V]
    ]
    [/state
      [cpu\_timers]
      [gpu\_timers]
      [int\_vals]
    ]
]
]
\end{forest}}
\par\end{centering}
\vspace{-5pt}
\caption{\label{fig:CLAMR-HDF5-file}The HDF5 file layout structure for CLAMR. }
\end{figure}
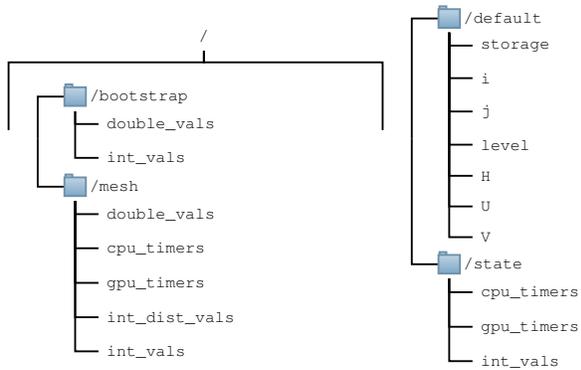

CLAMR (\textbf{C}ell-Based \textbf{A}daptive \textbf{M}esh \textbf{R}efinement)
is a testbed application for hybrid algorithm development using MPI
and OpenCL GPU code~\cite{Trujillo2011}. CLAMR does not use one output
file per process due to several issues: 
\begin{itemize}
\item File systems are limited in their ability to manage hundreds of thousands
of files;
\item In practice, managing hundreds of thousands of files is cumbersome
and error-prone;
\item Reading the data back using a different number of processes than the
analysis simulation requires redistribution and reshuffling of the
data, negating the advantage over more sophisticated collective I/O
strategies. 
\end{itemize}
Thus, the I/O strategy for CLAMR is to create one output file per
time step and to store each time step in an HDF5 file. Since HDF5
is a self-describing hierarchal file format, the ``metadata'' is
automatically handled by HDF5. Thus, using HDF5 significantly reduces
the internal bookkeeping and file construction required by CLAMR when
compared to using POSIX or MPI-IO. The HDF5 implementation organizes
the stored variables into datasets and uses groups to hold the datasets
themselves, Fig. \ref{fig:CLAMR-HDF5-file}.

Only a call to initialize DAOS was added to CLAMR's HDF5 implementation to utilize DAOS. This DAOS initialization requires a pool id which is passed to CLAMR by setting the environment variable \emph{pid}.
The pool id is obtained by starting the DAOS server:
\begin{tcolorbox}[colback=white,left=2pt,opacityframe=0.5]
\scriptsize
\begin{verbatim}
$ orterun -np 1 --report-uri ~/uri.txt daos_server -c 1
\end{verbatim}
\end{tcolorbox}where \texttt{c} is the number of DAOS server threads, and then creating
the pool and assigning the pool id to a shell environment variable:
\begin{tcolorbox}[colback=white,left=5pt,opacityframe=0.5]
\scriptsize
\begin{verbatim}
$ export pid=$(orterun -np 1 --ompi-server \
    file:~/uri.txt dmg create)
\end{verbatim}
\end{tcolorbox}
The first benchmark tested CLAMR for problem sizes ranging from
$2^{7}$ to $2^{10}$ grid points by powers of 2, with the number of processors
varying from 1 to 512 by powers of 2. For the initial run, a stripe count of four
and a stripe size of 4 MB was used. The time to write the checkpoint
files substantially increase as the number of processors is increased.

To gain a better understanding of what performance factors could be
tuned, the Lustre parameters were varied for a fixed problem size of
$2^{11}$ grid points. The Lustre parameters started at a default stripe count
of 1 with a stripe size of 1 MB, as well as the case of a stripe count
of 4 and stripe size of 1 MB. The results were compared to the
first case of a stripe count of 4 and stripe size of 4 MB. The obtained
results lead to the conclusion that the Lustre parameters had a slight
effect on performance, with the default stripe count of 1 and stripe size
of 1 MB resulted in the best throughput.

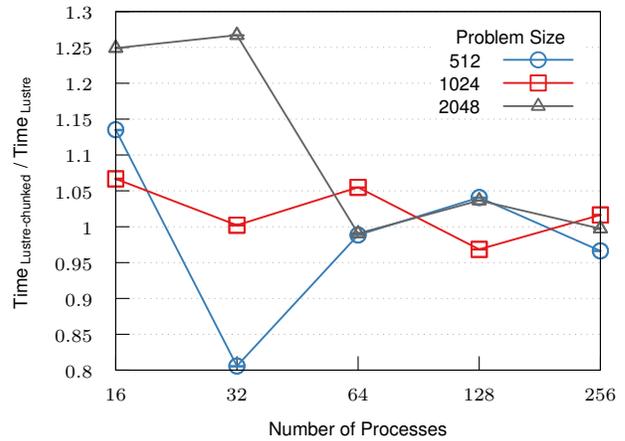
\begin{figure}
\begin{centering}
\begin{tikzpicture}[gnuplot]
%% generated with GNUPLOT 5.0p5 (Lua 5.3; terminal rev. 99, script rev. 100)
%% Thu 29 Jun 2017 05:23:30 PM CDT
\tikzset{every node/.append style={font={\sffamily \scriptsize}}}
\path (0.000,0.000) rectangle (8.500,6.125);
\gpcolor{color=gp lt color axes}
\gpsetlinetype{gp lt axes}
\gpsetdashtype{gp dt axes}
\gpsetlinewidth{0.50}
\draw[gp path] (1.504,0.985)--(7.947,0.985);
\gpcolor{color=gp lt color border}
\gpsetlinetype{gp lt border}
\gpsetdashtype{gp dt solid}
\gpsetlinewidth{1.00}
\draw[gp path] (1.504,0.985)--(1.684,0.985);
\node[gp node right] at (1.320,0.985) {$0.8$};
\gpcolor{color=gp lt color axes}
\gpsetlinetype{gp lt axes}
\gpsetdashtype{gp dt axes}
\gpsetlinewidth{0.50}
\draw[gp path] (1.504,1.462)--(7.947,1.462);
\gpcolor{color=gp lt color border}
\gpsetlinetype{gp lt border}
\gpsetdashtype{gp dt solid}
\gpsetlinewidth{1.00}
\draw[gp path] (1.504,1.462)--(1.684,1.462);
\node[gp node right] at (1.320,1.462) {$0.85$};
\gpcolor{color=gp lt color axes}
\gpsetlinetype{gp lt axes}
\gpsetdashtype{gp dt axes}
\gpsetlinewidth{0.50}
\draw[gp path] (1.504,1.939)--(7.947,1.939);
\gpcolor{color=gp lt color border}
\gpsetlinetype{gp lt border}
\gpsetdashtype{gp dt solid}
\gpsetlinewidth{1.00}
\draw[gp path] (1.504,1.939)--(1.684,1.939);
\node[gp node right] at (1.320,1.939) {$0.9$};
\gpcolor{color=gp lt color axes}
\gpsetlinetype{gp lt axes}
\gpsetdashtype{gp dt axes}
\gpsetlinewidth{0.50}
\draw[gp path] (1.504,2.416)--(7.947,2.416);
\gpcolor{color=gp lt color border}
\gpsetlinetype{gp lt border}
\gpsetdashtype{gp dt solid}
\gpsetlinewidth{1.00}
\draw[gp path] (1.504,2.416)--(1.684,2.416);
\node[gp node right] at (1.320,2.416) {$0.95$};
\gpcolor{color=gp lt color axes}
\gpsetlinetype{gp lt axes}
\gpsetdashtype{gp dt axes}
\gpsetlinewidth{0.50}
\draw[gp path] (1.504,2.893)--(7.947,2.893);
\gpcolor{color=gp lt color border}
\gpsetlinetype{gp lt border}
\gpsetdashtype{gp dt solid}
\gpsetlinewidth{1.00}
\draw[gp path] (1.504,2.893)--(1.684,2.893);
\node[gp node right] at (1.320,2.893) {$1$};
\gpcolor{color=gp lt color axes}
\gpsetlinetype{gp lt axes}
\gpsetdashtype{gp dt axes}
\gpsetlinewidth{0.50}
\draw[gp path] (1.504,3.371)--(7.947,3.371);
\gpcolor{color=gp lt color border}
\gpsetlinetype{gp lt border}
\gpsetdashtype{gp dt solid}
\gpsetlinewidth{1.00}
\draw[gp path] (1.504,3.371)--(1.684,3.371);
\node[gp node right] at (1.320,3.371) {$1.05$};
\gpcolor{color=gp lt color axes}
\gpsetlinetype{gp lt axes}
\gpsetdashtype{gp dt axes}
\gpsetlinewidth{0.50}
\draw[gp path] (1.504,3.848)--(7.947,3.848);
\gpcolor{color=gp lt color border}
\gpsetlinetype{gp lt border}
\gpsetdashtype{gp dt solid}
\gpsetlinewidth{1.00}
\draw[gp path] (1.504,3.848)--(1.684,3.848);
\node[gp node right] at (1.320,3.848) {$1.1$};
\gpcolor{color=gp lt color axes}
\gpsetlinetype{gp lt axes}
\gpsetdashtype{gp dt axes}
\gpsetlinewidth{0.50}
\draw[gp path] (1.504,4.325)--(7.947,4.325);
\gpcolor{color=gp lt color border}
\gpsetlinetype{gp lt border}
\gpsetdashtype{gp dt solid}
\gpsetlinewidth{1.00}
\draw[gp path] (1.504,4.325)--(1.684,4.325);
\node[gp node right] at (1.320,4.325) {$1.15$};
\gpcolor{color=gp lt color axes}
\gpsetlinetype{gp lt axes}
\gpsetdashtype{gp dt axes}
\gpsetlinewidth{0.50}
\draw[gp path] (1.504,4.802)--(5.743,4.802);
\draw[gp path] (7.763,4.802)--(7.947,4.802);
\gpcolor{color=gp lt color border}
\gpsetlinetype{gp lt border}
\gpsetdashtype{gp dt solid}
\gpsetlinewidth{1.00}
\draw[gp path] (1.504,4.802)--(1.684,4.802);
\node[gp node right] at (1.320,4.802) {$1.2$};
\gpcolor{color=gp lt color axes}
\gpsetlinetype{gp lt axes}
\gpsetdashtype{gp dt axes}
\gpsetlinewidth{0.50}
\draw[gp path] (1.504,5.279)--(5.743,5.279);
\draw[gp path] (7.763,5.279)--(7.947,5.279);
\gpcolor{color=gp lt color border}
\gpsetlinetype{gp lt border}
\gpsetdashtype{gp dt solid}
\gpsetlinewidth{1.00}
\draw[gp path] (1.504,5.279)--(1.684,5.279);
\node[gp node right] at (1.320,5.279) {$1.25$};
\gpcolor{color=gp lt color axes}
\gpsetlinetype{gp lt axes}
\gpsetdashtype{gp dt axes}
\gpsetlinewidth{0.50}
\draw[gp path] (1.504,5.756)--(7.947,5.756);
\gpcolor{color=gp lt color border}
\gpsetlinetype{gp lt border}
\gpsetdashtype{gp dt solid}
\gpsetlinewidth{1.00}
\draw[gp path] (1.504,5.756)--(1.684,5.756);
\node[gp node right] at (1.320,5.756) {$1.3$};
\draw[gp path] (1.504,0.985)--(1.504,1.165);
\node[gp node center] at (1.504,0.677) {$16$};
\draw[gp path] (3.115,0.985)--(3.115,1.165);
\node[gp node center] at (3.115,0.677) {$32$};
\draw[gp path] (4.726,0.985)--(4.726,1.165);
\node[gp node center] at (4.726,0.677) {$64$};
\draw[gp path] (6.336,0.985)--(6.336,1.165);
\node[gp node center] at (6.336,0.677) {$128$};
\draw[gp path] (7.947,0.985)--(7.947,1.165);
\node[gp node center] at (7.947,0.677) {$256$};
\draw[gp path] (1.504,5.756)--(1.504,0.985)--(7.947,0.985)--(7.947,5.756)--cycle;
\node[gp node center,rotate=-270] at (0.246,3.370) {Time\textsubscript{~Lustre-chunked} / Time\textsubscript{~Lustre}};
\node[gp node center] at (4.725,0.215) {Number of Processes};
\node[gp node center] at (6.753,5.422) {Problem Size};
\node[gp node right] at (6.479,5.114) {512};
\gpcolor{rgb color={0.216,0.494,0.722}}
\gpsetlinewidth{1.80}
\draw[gp path] (6.663,5.114)--(7.579,5.114);
\draw[gp path] (1.504,4.184)--(3.115,1.040)--(4.726,2.783)--(6.336,3.282)--(7.947,2.572);
\gpsetpointsize{7.20}
\gppoint{gp mark 6}{(1.504,4.184)}
\gppoint{gp mark 6}{(3.115,1.040)}
\gppoint{gp mark 6}{(4.726,2.783)}
\gppoint{gp mark 6}{(6.336,3.282)}
\gppoint{gp mark 6}{(7.947,2.572)}
\gppoint{gp mark 6}{(7.121,5.114)}
\draw[gp path] (1.504,4.184)--(1.504,4.175);
\draw[gp path] (1.504,4.184)--(1.594,4.184);
\draw[gp path] (1.504,4.175)--(1.594,4.175);
\draw[gp path] (3.115,1.033)--(3.115,1.040);
\draw[gp path] (3.025,1.033)--(3.205,1.033);
\draw[gp path] (3.025,1.040)--(3.205,1.040);
\draw[gp path] (4.726,2.781)--(4.726,2.790);
\draw[gp path] (4.636,2.781)--(4.816,2.781);
\draw[gp path] (4.636,2.790)--(4.816,2.790);
\draw[gp path] (6.336,3.280)--(6.336,3.283);
\draw[gp path] (6.246,3.280)--(6.426,3.280);
\draw[gp path] (6.246,3.283)--(6.426,3.283);
\draw[gp path] (7.947,2.570)--(7.947,2.573);
\draw[gp path] (7.857,2.570)--(7.947,2.570);
\draw[gp path] (7.857,2.573)--(7.947,2.573);
\gppoint{gp mark 6}{(1.504,4.184)}
\gppoint{gp mark 6}{(3.115,1.040)}
\gppoint{gp mark 6}{(4.726,2.783)}
\gppoint{gp mark 6}{(6.336,3.282)}
\gppoint{gp mark 6}{(7.947,2.572)}
\gpcolor{color=gp lt color border}
\node[gp node right] at (6.479,4.806) {1024};
\gpcolor{rgb color={0.894,0.102,0.110}}
\draw[gp path] (6.663,4.806)--(7.579,4.806);
\draw[gp path] (1.504,3.530)--(3.115,2.913)--(4.726,3.417)--(6.336,2.592)--(7.947,3.050);
\gppoint{gp mark 4}{(1.504,3.530)}
\gppoint{gp mark 4}{(3.115,2.913)}
\gppoint{gp mark 4}{(4.726,3.417)}
\gppoint{gp mark 4}{(6.336,2.592)}
\gppoint{gp mark 4}{(7.947,3.050)}
\gppoint{gp mark 4}{(7.121,4.806)}
\draw[gp path] (1.504,3.523)--(1.504,3.535);
\draw[gp path] (1.504,3.523)--(1.594,3.523);
\draw[gp path] (1.504,3.535)--(1.594,3.535);
\draw[gp path] (3.115,2.909)--(3.115,2.925);
\draw[gp path] (3.025,2.909)--(3.205,2.909);
\draw[gp path] (3.025,2.925)--(3.205,2.925);
\draw[gp path] (4.636,3.415)--(4.816,3.415);
\draw[gp path] (4.636,3.415)--(4.816,3.415);
\draw[gp path] (6.336,2.589)--(6.336,2.595);
\draw[gp path] (6.246,2.589)--(6.426,2.589);
\draw[gp path] (6.246,2.595)--(6.426,2.595);
\draw[gp path] (7.947,3.048)--(7.947,3.052);
\draw[gp path] (7.857,3.048)--(7.947,3.048);
\draw[gp path] (7.857,3.052)--(7.947,3.052);
\gppoint{gp mark 4}{(1.504,3.530)}
\gppoint{gp mark 4}{(3.115,2.913)}
\gppoint{gp mark 4}{(4.726,3.417)}
\gppoint{gp mark 4}{(6.336,2.592)}
\gppoint{gp mark 4}{(7.947,3.050)}
\gpcolor{color=gp lt color border}
\node[gp node right] at (6.479,4.498) {2048};
\gpcolor{rgb color={0.400,0.400,0.400}}
\draw[gp path] (6.663,4.498)--(7.579,4.498);
\draw[gp path] (1.504,5.270)--(3.115,5.442)--(4.726,2.803)--(6.336,3.242)--(7.947,2.868);
\gppoint{gp mark 8}{(1.504,5.270)}
\gppoint{gp mark 8}{(3.115,5.442)}
\gppoint{gp mark 8}{(4.726,2.803)}
\gppoint{gp mark 8}{(6.336,3.242)}
\gppoint{gp mark 8}{(7.947,2.868)}
\gppoint{gp mark 8}{(7.121,4.498)}
\draw[gp path] (1.504,5.269)--(1.504,5.273);
\draw[gp path] (1.504,5.269)--(1.594,5.269);
\draw[gp path] (1.504,5.273)--(1.594,5.273);
\draw[gp path] (3.115,5.441)--(3.115,5.439);
\draw[gp path] (3.025,5.441)--(3.205,5.441);
\draw[gp path] (3.025,5.439)--(3.205,5.439);
\draw[gp path] (4.726,2.801)--(4.726,2.809);
\draw[gp path] (4.636,2.801)--(4.816,2.801);
\draw[gp path] (4.636,2.809)--(4.816,2.809);
\draw[gp path] (6.336,3.241)--(6.336,3.244);
\draw[gp path] (6.246,3.241)--(6.426,3.241);
\draw[gp path] (6.246,3.244)--(6.426,3.244);
\draw[gp path] (7.947,2.862)--(7.947,2.875);
\draw[gp path] (7.857,2.862)--(7.947,2.862);
\draw[gp path] (7.857,2.875)--(7.947,2.875);
\gppoint{gp mark 8}{(1.504,5.270)}
\gppoint{gp mark 8}{(3.115,5.442)}
\gppoint{gp mark 8}{(4.726,2.803)}
\gppoint{gp mark 8}{(6.336,3.242)}
\gppoint{gp mark 8}{(7.947,2.868)}
\gpcolor{color=gp lt color border}
\gpsetlinewidth{1.00}
\draw[gp path] (1.504,5.756)--(1.504,0.985)--(7.947,0.985)--(7.947,5.756)--cycle;
%% coordinates of the plot area
\gpdefrectangularnode{gp plot 1}{\pgfpoint{1.504cm}{0.985cm}}{\pgfpoint{7.947cm}{5.756cm}}
\end{tikzpicture}
%% gnuplot variables
\par\end{centering}
\vspace{-15pt}
\caption{\label{fig:Comparison-of-Lustre-2}
Comparison of CLAMR's Lustre write performance between HDF5 unchunked I/O and HDF5 chunked I/O using the optimal chunk size for the given problem size. Values greater than one represents chunked CLAMR I/O performing worse than unchunked CLAMR I/O.}
\end{figure}

Finally, different HDF5 chunking layout parameters using
the default Lustre settings of a stripe count of 1 and stripe size
of 1 MB were investigated. A variety of chunk sizes from $2^{8}$ to
$2^{18}$ by powers of 2 were tested across different problem sizes
to obtain the optimal chunk size for the given problem size. 
The results show an improvement in CLAMR's performance when using the optimal chunk size in comparison to the  nonchunked I/O performance,
Fig. \ref{fig:Comparison-of-Lustre-2}.

These results show that chunking had little positive effect on the
performance of CLAMR checkpoint writing, with the result being marginally
faster in a few cases and on par with or significantly worse than unchunked
I/O performance in the other cases. When the chunk size of $8192$ was used 
with $32$ processors on the $512$ problem size, a performance increase of
$20$\% is observed, however no chunk size was able to replicate this
performance improvement for the other sets of parameters.

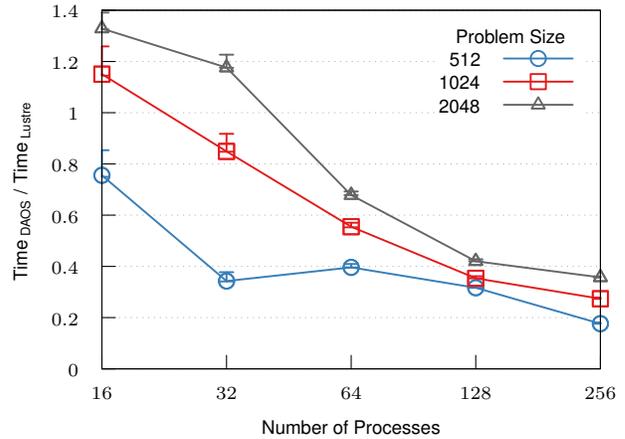
\begin{figure}
\begin{centering}
\begin{tikzpicture}[gnuplot]
%% generated with GNUPLOT 5.0p5 (Lua 5.3; terminal rev. 99, script rev. 100)
%% Thu 29 Jun 2017 01:59:32 PM CDT
\tikzset{every node/.append style={font={\sffamily \scriptsize}}}
\path (0.000,0.000) rectangle (8.500,6.125);
\gpcolor{color=gp lt color axes}
\gpsetlinetype{gp lt axes}
\gpsetdashtype{gp dt axes}
\gpsetlinewidth{0.50}
\draw[gp path] (1.320,0.985)--(7.947,0.985);
\gpcolor{color=gp lt color border}
\gpsetlinetype{gp lt border}
\gpsetdashtype{gp dt solid}
\gpsetlinewidth{1.00}
\draw[gp path] (1.320,0.985)--(1.500,0.985);
\node[gp node right] at (1.136,0.985) {$0$};
\gpcolor{color=gp lt color axes}
\gpsetlinetype{gp lt axes}
\gpsetdashtype{gp dt axes}
\gpsetlinewidth{0.50}
\draw[gp path] (1.320,1.667)--(7.947,1.667);
\gpcolor{color=gp lt color border}
\gpsetlinetype{gp lt border}
\gpsetdashtype{gp dt solid}
\gpsetlinewidth{1.00}
\draw[gp path] (1.320,1.667)--(1.500,1.667);
\node[gp node right] at (1.136,1.667) {$0.2$};
\gpcolor{color=gp lt color axes}
\gpsetlinetype{gp lt axes}
\gpsetdashtype{gp dt axes}
\gpsetlinewidth{0.50}
\draw[gp path] (1.320,2.348)--(7.947,2.348);
\gpcolor{color=gp lt color border}
\gpsetlinetype{gp lt border}
\gpsetdashtype{gp dt solid}
\gpsetlinewidth{1.00}
\draw[gp path] (1.320,2.348)--(1.500,2.348);
\node[gp node right] at (1.136,2.348) {$0.4$};
\gpcolor{color=gp lt color axes}
\gpsetlinetype{gp lt axes}
\gpsetdashtype{gp dt axes}
\gpsetlinewidth{0.50}
\draw[gp path] (1.320,3.030)--(7.947,3.030);
\gpcolor{color=gp lt color border}
\gpsetlinetype{gp lt border}
\gpsetdashtype{gp dt solid}
\gpsetlinewidth{1.00}
\draw[gp path] (1.320,3.030)--(1.500,3.030);
\node[gp node right] at (1.136,3.030) {$0.6$};
\gpcolor{color=gp lt color axes}
\gpsetlinetype{gp lt axes}
\gpsetdashtype{gp dt axes}
\gpsetlinewidth{0.50}
\draw[gp path] (1.320,3.711)--(7.947,3.711);
\gpcolor{color=gp lt color border}
\gpsetlinetype{gp lt border}
\gpsetdashtype{gp dt solid}
\gpsetlinewidth{1.00}
\draw[gp path] (1.320,3.711)--(1.500,3.711);
\node[gp node right] at (1.136,3.711) {$0.8$};
\gpcolor{color=gp lt color axes}
\gpsetlinetype{gp lt axes}
\gpsetdashtype{gp dt axes}
\gpsetlinewidth{0.50}
\draw[gp path] (1.320,4.393)--(5.743,4.393);
\draw[gp path] (7.763,4.393)--(7.947,4.393);
\gpcolor{color=gp lt color border}
\gpsetlinetype{gp lt border}
\gpsetdashtype{gp dt solid}
\gpsetlinewidth{1.00}
\draw[gp path] (1.320,4.393)--(1.500,4.393);
\node[gp node right] at (1.136,4.393) {$1$};
\gpcolor{color=gp lt color axes}
\gpsetlinetype{gp lt axes}
\gpsetdashtype{gp dt axes}
\gpsetlinewidth{0.50}
\draw[gp path] (1.320,5.074)--(5.743,5.074);
\draw[gp path] (7.763,5.074)--(7.947,5.074);
\gpcolor{color=gp lt color border}
\gpsetlinetype{gp lt border}
\gpsetdashtype{gp dt solid}
\gpsetlinewidth{1.00}
\draw[gp path] (1.320,5.074)--(1.500,5.074);
\node[gp node right] at (1.136,5.074) {$1.2$};
\gpcolor{color=gp lt color axes}
\gpsetlinetype{gp lt axes}
\gpsetdashtype{gp dt axes}
\gpsetlinewidth{0.50}
\draw[gp path] (1.320,5.756)--(7.947,5.756);
\gpcolor{color=gp lt color border}
\gpsetlinetype{gp lt border}
\gpsetdashtype{gp dt solid}
\gpsetlinewidth{1.00}
\draw[gp path] (1.320,5.756)--(1.500,5.756);
\node[gp node right] at (1.136,5.756) {$1.4$};
\draw[gp path] (1.320,0.985)--(1.320,1.165);
\node[gp node center] at (1.320,0.677) {$16$};
\draw[gp path] (2.977,0.985)--(2.977,1.165);
\node[gp node center] at (2.977,0.677) {$32$};
\draw[gp path] (4.634,0.985)--(4.634,1.165);
\node[gp node center] at (4.634,0.677) {$64$};
\draw[gp path] (6.290,0.985)--(6.290,1.165);
\node[gp node center] at (6.290,0.677) {$128$};
\draw[gp path] (7.947,0.985)--(7.947,1.165);
\node[gp node center] at (7.947,0.677) {$256$};
\draw[gp path] (1.320,5.756)--(1.320,0.985)--(7.947,0.985)--(7.947,5.756)--cycle;
\node[gp node center,rotate=-270] at (0.246,3.370) {Time\textsubscript{~DAOS} / Time\textsubscript{~Lustre}};
\node[gp node center] at (4.633,0.215) {Number of Processes};
\node[gp node center] at (6.753,5.422) {Problem Size};
\node[gp node right] at (6.479,5.114) {512};
\gpcolor{rgb color={0.216,0.494,0.722}}
\gpsetlinewidth{1.80}
\draw[gp path] (6.663,5.114)--(7.579,5.114);
\draw[gp path] (1.320,3.560)--(2.977,2.153)--(4.634,2.335)--(6.290,2.064)--(7.947,1.586);
\gpsetpointsize{7.20}
\gppoint{gp mark 6}{(1.320,3.560)}
\gppoint{gp mark 6}{(2.977,2.153)}
\gppoint{gp mark 6}{(4.634,2.335)}
\gppoint{gp mark 6}{(6.290,2.064)}
\gppoint{gp mark 6}{(7.947,1.586)}
\gppoint{gp mark 6}{(7.121,5.114)}
\draw[gp path] (1.320,3.540)--(1.320,3.894);
\draw[gp path] (1.320,3.540)--(1.410,3.540);
\draw[gp path] (1.320,3.894)--(1.410,3.894);
\draw[gp path] (2.977,2.148)--(2.977,2.271);
\draw[gp path] (2.887,2.148)--(3.067,2.148);
\draw[gp path] (2.887,2.271)--(3.067,2.271);
\draw[gp path] (4.634,2.334)--(4.634,2.380);
\draw[gp path] (4.544,2.334)--(4.724,2.334);
\draw[gp path] (4.544,2.380)--(4.724,2.380);
\draw[gp path] (6.290,2.063)--(6.290,2.090);
\draw[gp path] (6.200,2.063)--(6.380,2.063);
\draw[gp path] (6.200,2.090)--(6.380,2.090);
\draw[gp path] (7.947,1.585)--(7.947,1.601);
\draw[gp path] (7.857,1.585)--(7.947,1.585);
\draw[gp path] (7.857,1.601)--(7.947,1.601);
\gppoint{gp mark 6}{(1.320,3.560)}
\gppoint{gp mark 6}{(2.977,2.153)}
\gppoint{gp mark 6}{(4.634,2.335)}
\gppoint{gp mark 6}{(6.290,2.064)}
\gppoint{gp mark 6}{(7.947,1.586)}
\gpcolor{color=gp lt color border}
\node[gp node right] at (6.479,4.806) {1024};
\gpcolor{rgb color={0.894,0.102,0.110}}
\draw[gp path] (6.663,4.806)--(7.579,4.806);
\draw[gp path] (1.320,4.906)--(2.977,3.878)--(4.634,2.876)--(6.290,2.190)--(7.947,1.918);
\gppoint{gp mark 4}{(1.320,4.906)}
\gppoint{gp mark 4}{(2.977,3.878)}
\gppoint{gp mark 4}{(4.634,2.876)}
\gppoint{gp mark 4}{(6.290,2.190)}
\gppoint{gp mark 4}{(7.947,1.918)}
\gppoint{gp mark 4}{(7.121,4.806)}
\draw[gp path] (1.320,4.886)--(1.320,5.275);
\draw[gp path] (1.320,4.886)--(1.410,4.886);
\draw[gp path] (1.320,5.275)--(1.410,5.275);
\draw[gp path] (2.977,3.874)--(2.977,4.112);
\draw[gp path] (2.887,3.874)--(3.067,3.874);
\draw[gp path] (2.887,4.112)--(3.067,4.112);
\draw[gp path] (4.634,2.868)--(4.634,2.929);
\draw[gp path] (4.544,2.868)--(4.724,2.868);
\draw[gp path] (4.544,2.929)--(4.724,2.929);
\draw[gp path] (6.290,2.189)--(6.290,2.217);
\draw[gp path] (6.200,2.189)--(6.380,2.189);
\draw[gp path] (6.200,2.217)--(6.380,2.217);
\draw[gp path] (7.947,1.916)--(7.947,1.933);
\draw[gp path] (7.857,1.916)--(7.947,1.916);
\draw[gp path] (7.857,1.933)--(7.947,1.933);
\gppoint{gp mark 4}{(1.320,4.906)}
\gppoint{gp mark 4}{(2.977,3.878)}
\gppoint{gp mark 4}{(4.634,2.876)}
\gppoint{gp mark 4}{(6.290,2.190)}
\gppoint{gp mark 4}{(7.947,1.918)}
\gpcolor{color=gp lt color border}
\node[gp node right] at (6.479,4.498) {2048};
\gpcolor{rgb color={0.400,0.400,0.400}}
\draw[gp path] (6.663,4.498)--(7.579,4.498);
\draw[gp path] (1.320,5.513)--(2.977,4.993)--(4.634,3.297)--(6.290,2.417)--(7.947,2.202);
\gppoint{gp mark 8}{(1.320,5.513)}
\gppoint{gp mark 8}{(2.977,4.993)}
\gppoint{gp mark 8}{(4.634,3.297)}
\gppoint{gp mark 8}{(6.290,2.417)}
\gppoint{gp mark 8}{(7.947,2.202)}
\gppoint{gp mark 8}{(7.121,4.498)}
\draw[gp path] (1.320,5.501)--(1.320,5.725);
\draw[gp path] (1.320,5.501)--(1.410,5.501);
\draw[gp path] (1.320,5.725)--(1.410,5.725);
\draw[gp path] (2.977,4.990)--(2.977,5.165);
\draw[gp path] (2.887,4.990)--(3.067,4.990);
\draw[gp path] (2.887,5.165)--(3.067,5.165);
\draw[gp path] (4.634,3.297)--(4.634,3.345);
\draw[gp path] (4.544,3.297)--(4.724,3.297);
\draw[gp path] (4.544,3.345)--(4.724,3.345);
\draw[gp path] (6.290,2.416)--(6.290,2.443);
\draw[gp path] (6.200,2.416)--(6.380,2.416);
\draw[gp path] (6.200,2.443)--(6.380,2.443);
\draw[gp path] (7.947,2.200)--(7.947,2.215);
\draw[gp path] (7.857,2.200)--(7.947,2.200);
\draw[gp path] (7.857,2.215)--(7.947,2.215);
\gppoint{gp mark 8}{(1.320,5.513)}
\gppoint{gp mark 8}{(2.977,4.993)}
\gppoint{gp mark 8}{(4.634,3.297)}
\gppoint{gp mark 8}{(6.290,2.417)}
\gppoint{gp mark 8}{(7.947,2.202)}
\gpcolor{color=gp lt color border}
\gpsetlinewidth{1.00}
\draw[gp path] (1.320,5.756)--(1.320,0.985)--(7.947,0.985)--(7.947,5.756)--cycle;
%% coordinates of the plot area
\gpdefrectangularnode{gp plot 1}{\pgfpoint{1.320cm}{0.985cm}}{\pgfpoint{7.947cm}{5.756cm}}
\end{tikzpicture}
%% gnuplot variables
\par\end{centering}
\vspace{-15pt}
\caption{\label{fig:CLAMR_daos_lustre}Comparison of unchunked write performance
of CLAMR on DAOS/Lustre. Lustre parameters are a stripe count of one and stripe size
of 1MB and DAOS used one DAOS server. Values above one represent unchunked
CLAMR I/O on DAOS performing worse than unchunked CLAMR I/O .}
\end{figure}

\begin{figure}
\begin{centering}
\begin{tikzpicture}[gnuplot]
%% generated with GNUPLOT 5.0p5 (Lua 5.3; terminal rev. 99, script rev. 100)
%% Thu 29 Jun 2017 01:59:32 PM CDT
\tikzset{every node/.append style={font={\sffamily \scriptsize}}}
\path (0.000,0.000) rectangle (8.500,6.125);
\gpcolor{color=gp lt color axes}
\gpsetlinetype{gp lt axes}
\gpsetdashtype{gp dt axes}
\gpsetlinewidth{0.50}
\draw[gp path] (1.320,0.985)--(7.947,0.985);
\gpcolor{color=gp lt color border}
\gpsetlinetype{gp lt border}
\gpsetdashtype{gp dt solid}
\gpsetlinewidth{1.00}
\draw[gp path] (1.320,0.985)--(1.500,0.985);
\node[gp node right] at (1.136,0.985) {$0.4$};
\gpcolor{color=gp lt color axes}
\gpsetlinetype{gp lt axes}
\gpsetdashtype{gp dt axes}
\gpsetlinewidth{0.50}
\draw[gp path] (1.320,1.667)--(7.947,1.667);
\gpcolor{color=gp lt color border}
\gpsetlinetype{gp lt border}
\gpsetdashtype{gp dt solid}
\gpsetlinewidth{1.00}
\draw[gp path] (1.320,1.667)--(1.500,1.667);
\node[gp node right] at (1.136,1.667) {$0.6$};
\gpcolor{color=gp lt color axes}
\gpsetlinetype{gp lt axes}
\gpsetdashtype{gp dt axes}
\gpsetlinewidth{0.50}
\draw[gp path] (1.320,2.348)--(7.947,2.348);
\gpcolor{color=gp lt color border}
\gpsetlinetype{gp lt border}
\gpsetdashtype{gp dt solid}
\gpsetlinewidth{1.00}
\draw[gp path] (1.320,2.348)--(1.500,2.348);
\node[gp node right] at (1.136,2.348) {$0.8$};
\gpcolor{color=gp lt color axes}
\gpsetlinetype{gp lt axes}
\gpsetdashtype{gp dt axes}
\gpsetlinewidth{0.50}
\draw[gp path] (1.320,3.030)--(7.947,3.030);
\gpcolor{color=gp lt color border}
\gpsetlinetype{gp lt border}
\gpsetdashtype{gp dt solid}
\gpsetlinewidth{1.00}
\draw[gp path] (1.320,3.030)--(1.500,3.030);
\node[gp node right] at (1.136,3.030) {$1$};
\gpcolor{color=gp lt color axes}
\gpsetlinetype{gp lt axes}
\gpsetdashtype{gp dt axes}
\gpsetlinewidth{0.50}
\draw[gp path] (1.320,3.711)--(7.947,3.711);
\gpcolor{color=gp lt color border}
\gpsetlinetype{gp lt border}
\gpsetdashtype{gp dt solid}
\gpsetlinewidth{1.00}
\draw[gp path] (1.320,3.711)--(1.500,3.711);
\node[gp node right] at (1.136,3.711) {$1.2$};
\gpcolor{color=gp lt color axes}
\gpsetlinetype{gp lt axes}
\gpsetdashtype{gp dt axes}
\gpsetlinewidth{0.50}
\draw[gp path] (1.320,4.393)--(5.743,4.393);
\draw[gp path] (7.763,4.393)--(7.947,4.393);
\gpcolor{color=gp lt color border}
\gpsetlinetype{gp lt border}
\gpsetdashtype{gp dt solid}
\gpsetlinewidth{1.00}
\draw[gp path] (1.320,4.393)--(1.500,4.393);
\node[gp node right] at (1.136,4.393) {$1.4$};
\gpcolor{color=gp lt color axes}
\gpsetlinetype{gp lt axes}
\gpsetdashtype{gp dt axes}
\gpsetlinewidth{0.50}
\draw[gp path] (1.320,5.074)--(5.743,5.074);
\draw[gp path] (7.763,5.074)--(7.947,5.074);
\gpcolor{color=gp lt color border}
\gpsetlinetype{gp lt border}
\gpsetdashtype{gp dt solid}
\gpsetlinewidth{1.00}
\draw[gp path] (1.320,5.074)--(1.500,5.074);
\node[gp node right] at (1.136,5.074) {$1.6$};
\gpcolor{color=gp lt color axes}
\gpsetlinetype{gp lt axes}
\gpsetdashtype{gp dt axes}
\gpsetlinewidth{0.50}
\draw[gp path] (1.320,5.756)--(7.947,5.756);
\gpcolor{color=gp lt color border}
\gpsetlinetype{gp lt border}
\gpsetdashtype{gp dt solid}
\gpsetlinewidth{1.00}
\draw[gp path] (1.320,5.756)--(1.500,5.756);
\node[gp node right] at (1.136,5.756) {$1.8$};
\draw[gp path] (1.320,0.985)--(1.320,1.165);
\node[gp node center] at (1.320,0.677) {$16$};
\draw[gp path] (2.977,0.985)--(2.977,1.165);
\node[gp node center] at (2.977,0.677) {$32$};
\draw[gp path] (4.634,0.985)--(4.634,1.165);
\node[gp node center] at (4.634,0.677) {$64$};
\draw[gp path] (6.290,0.985)--(6.290,1.165);
\node[gp node center] at (6.290,0.677) {$128$};
\draw[gp path] (7.947,0.985)--(7.947,1.165);
\node[gp node center] at (7.947,0.677) {$256$};
\draw[gp path] (1.320,5.756)--(1.320,0.985)--(7.947,0.985)--(7.947,5.756)--cycle;
\node[gp node center,rotate=-270] at (0.246,3.370) {Time\textsubscript{~DAOS-Chunked} / Time\textsubscript{~DAOS}};
\node[gp node center] at (4.633,0.215) {Number of Processes};
\node[gp node center] at (6.753,5.422) {Problem Size};
\node[gp node right] at (6.479,5.114) {512};
\gpcolor{rgb color={0.216,0.494,0.722}}
\gpsetlinewidth{1.80}
\draw[gp path] (6.663,5.114)--(7.579,5.114);
\draw[gp path] (1.320,3.671)--(2.977,5.270)--(4.634,1.618)--(6.290,1.488)--(7.947,2.934);
\gpsetpointsize{7.20}
\gppoint{gp mark 6}{(1.320,3.671)}
\gppoint{gp mark 6}{(2.977,5.270)}
\gppoint{gp mark 6}{(4.634,1.618)}
\gppoint{gp mark 6}{(6.290,1.488)}
\gppoint{gp mark 6}{(7.947,2.934)}
\gppoint{gp mark 6}{(7.121,5.114)}
\draw[gp path] (1.320,3.687)--(1.320,3.456);
\draw[gp path] (1.320,3.687)--(1.410,3.687);
\draw[gp path] (1.320,3.456)--(1.410,3.456);
\draw[gp path] (2.977,5.281)--(2.977,4.965);
\draw[gp path] (2.887,5.281)--(3.067,5.281);
\draw[gp path] (2.887,4.965)--(3.067,4.965);
\draw[gp path] (4.634,1.617)--(4.634,1.633);
\draw[gp path] (4.544,1.617)--(4.724,1.617);
\draw[gp path] (4.544,1.633)--(4.724,1.633);
\draw[gp path] (6.290,1.485)--(6.290,1.499);
\draw[gp path] (6.200,1.485)--(6.380,1.485);
\draw[gp path] (6.200,1.499)--(6.380,1.499);
\draw[gp path] (7.947,2.934)--(7.947,2.917);
\draw[gp path] (7.857,2.934)--(7.947,2.934);
\draw[gp path] (7.857,2.917)--(7.947,2.917);
\gppoint{gp mark 6}{(1.320,3.671)}
\gppoint{gp mark 6}{(2.977,5.270)}
\gppoint{gp mark 6}{(4.634,1.618)}
\gppoint{gp mark 6}{(6.290,1.488)}
\gppoint{gp mark 6}{(7.947,2.934)}
\gpcolor{color=gp lt color border}
\node[gp node right] at (6.479,4.806) {1024};
\gpcolor{rgb color={0.894,0.102,0.110}}
\draw[gp path] (6.663,4.806)--(7.579,4.806);
\draw[gp path] (1.320,3.515)--(2.977,4.828)--(4.634,1.554)--(6.290,1.643)--(7.947,1.705);
\gppoint{gp mark 4}{(1.320,3.515)}
\gppoint{gp mark 4}{(2.977,4.828)}
\gppoint{gp mark 4}{(4.634,1.554)}
\gppoint{gp mark 4}{(6.290,1.643)}
\gppoint{gp mark 4}{(7.947,1.705)}
\gppoint{gp mark 4}{(7.121,4.806)}
\draw[gp path] (1.320,3.526)--(1.320,3.365);
\draw[gp path] (1.320,3.526)--(1.410,3.526);
\draw[gp path] (1.320,3.365)--(1.410,3.365);
\draw[gp path] (2.977,4.829)--(2.977,4.608);
\draw[gp path] (2.887,4.829)--(3.067,4.829);
\draw[gp path] (2.887,4.608)--(3.067,4.608);
\draw[gp path] (4.634,1.561)--(4.634,1.567);
\draw[gp path] (4.544,1.561)--(4.724,1.561);
\draw[gp path] (4.544,1.567)--(4.724,1.567);
\draw[gp path] (6.290,1.641)--(6.290,1.673);
\draw[gp path] (6.200,1.641)--(6.380,1.641);
\draw[gp path] (6.200,1.673)--(6.380,1.673);
\draw[gp path] (7.947,1.703)--(7.947,1.706);
\draw[gp path] (7.857,1.703)--(7.947,1.703);
\draw[gp path] (7.857,1.706)--(7.947,1.706);
\gppoint{gp mark 4}{(1.320,3.515)}
\gppoint{gp mark 4}{(2.977,4.828)}
\gppoint{gp mark 4}{(4.634,1.554)}
\gppoint{gp mark 4}{(6.290,1.643)}
\gppoint{gp mark 4}{(7.947,1.705)}
\gpcolor{color=gp lt color border}
\node[gp node right] at (6.479,4.498) {2048};
\gpcolor{rgb color={0.400,0.400,0.400}}
\draw[gp path] (6.663,4.498)--(7.579,4.498);
\draw[gp path] (1.320,3.762)--(2.977,3.656)--(4.634,1.377)--(6.290,1.482)--(7.947,1.910);
\gppoint{gp mark 8}{(1.320,3.762)}
\gppoint{gp mark 8}{(2.977,3.656)}
\gppoint{gp mark 8}{(4.634,1.377)}
\gppoint{gp mark 8}{(6.290,1.482)}
\gppoint{gp mark 8}{(7.947,1.910)}
\gppoint{gp mark 8}{(7.121,4.498)}
\draw[gp path] (1.320,3.766)--(1.320,3.673);
\draw[gp path] (1.320,3.766)--(1.410,3.766);
\draw[gp path] (1.320,3.673)--(1.410,3.673);
\draw[gp path] (2.977,3.657)--(2.977,3.582);
\draw[gp path] (2.887,3.657)--(3.067,3.657);
\draw[gp path] (2.887,3.582)--(3.067,3.582);
\draw[gp path] (4.634,1.376)--(4.634,1.388);
\draw[gp path] (4.544,1.376)--(4.724,1.376);
\draw[gp path] (4.544,1.388)--(4.724,1.388);
\draw[gp path] (6.290,1.480)--(6.290,1.494);
\draw[gp path] (6.200,1.480)--(6.380,1.480);
\draw[gp path] (6.200,1.494)--(6.380,1.494);
\draw[gp path] (7.947,1.911)--(7.947,1.914);
\draw[gp path] (7.857,1.911)--(7.947,1.911);
\draw[gp path] (7.857,1.914)--(7.947,1.914);
\gppoint{gp mark 8}{(1.320,3.762)}
\gppoint{gp mark 8}{(2.977,3.656)}
\gppoint{gp mark 8}{(4.634,1.377)}
\gppoint{gp mark 8}{(6.290,1.482)}
\gppoint{gp mark 8}{(7.947,1.910)}
\gpcolor{color=gp lt color border}
\gpsetlinewidth{1.00}
\draw[gp path] (1.320,5.756)--(1.320,0.985)--(7.947,0.985)--(7.947,5.756)--cycle;
%% coordinates of the plot area
\gpdefrectangularnode{gp plot 1}{\pgfpoint{1.320cm}{0.985cm}}{\pgfpoint{7.947cm}{5.756cm}}
\end{tikzpicture}
%% gnuplot variables
\par\end{centering}
\vspace{-15pt}
\caption{\label{fig:CLAMR_daos_daoschunked}
Comparison of chunked and unchunked CLAMR
write performance on DAOS. Unchunked CLAMR I/O used one DAOS server; Chunked CLAMR I/O used eight DAOS servers across eight different nodes. Values above one represent chunked
CLAMR I/O  performing worse than unchunked CLAMR I/O.}
\end{figure}
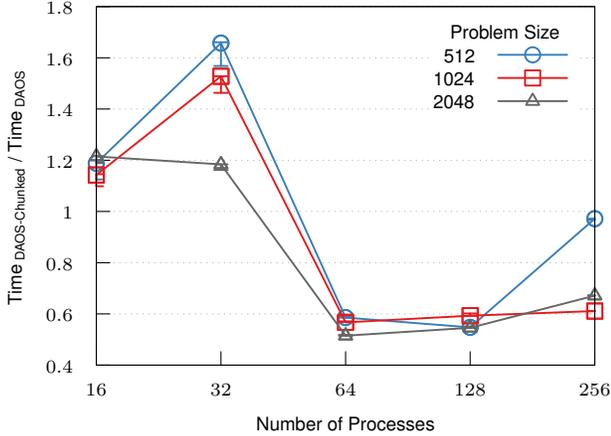

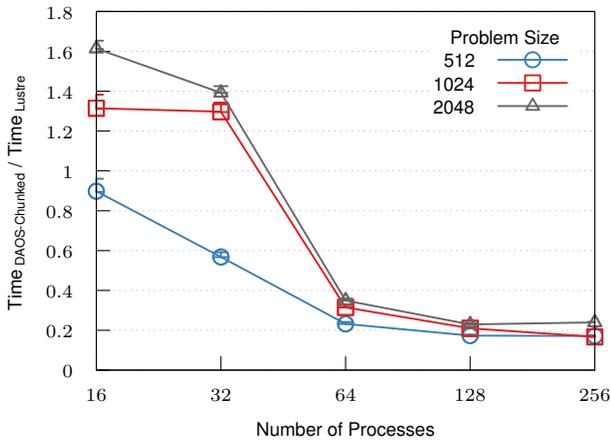
\begin{figure}
\begin{centering}
\begin{tikzpicture}[gnuplot]
%% generated with GNUPLOT 5.0p5 (Lua 5.3; terminal rev. 99, script rev. 100)
%% Thu 29 Jun 2017 01:59:32 PM CDT
\tikzset{every node/.append style={font={\sffamily \scriptsize}}}
\path (0.000,0.000) rectangle (8.500,6.125);
\gpcolor{color=gp lt color axes}
\gpsetlinetype{gp lt axes}
\gpsetdashtype{gp dt axes}
\gpsetlinewidth{0.50}
\draw[gp path] (1.320,0.985)--(7.947,0.985);
\gpcolor{color=gp lt color border}
\gpsetlinetype{gp lt border}
\gpsetdashtype{gp dt solid}
\gpsetlinewidth{1.00}
\draw[gp path] (1.320,0.985)--(1.500,0.985);
\node[gp node right] at (1.136,0.985) {$0$};
\gpcolor{color=gp lt color axes}
\gpsetlinetype{gp lt axes}
\gpsetdashtype{gp dt axes}
\gpsetlinewidth{0.50}
\draw[gp path] (1.320,1.515)--(7.947,1.515);
\gpcolor{color=gp lt color border}
\gpsetlinetype{gp lt border}
\gpsetdashtype{gp dt solid}
\gpsetlinewidth{1.00}
\draw[gp path] (1.320,1.515)--(1.500,1.515);
\node[gp node right] at (1.136,1.515) {$0.2$};
\gpcolor{color=gp lt color axes}
\gpsetlinetype{gp lt axes}
\gpsetdashtype{gp dt axes}
\gpsetlinewidth{0.50}
\draw[gp path] (1.320,2.045)--(7.947,2.045);
\gpcolor{color=gp lt color border}
\gpsetlinetype{gp lt border}
\gpsetdashtype{gp dt solid}
\gpsetlinewidth{1.00}
\draw[gp path] (1.320,2.045)--(1.500,2.045);
\node[gp node right] at (1.136,2.045) {$0.4$};
\gpcolor{color=gp lt color axes}
\gpsetlinetype{gp lt axes}
\gpsetdashtype{gp dt axes}
\gpsetlinewidth{0.50}
\draw[gp path] (1.320,2.575)--(7.947,2.575);
\gpcolor{color=gp lt color border}
\gpsetlinetype{gp lt border}
\gpsetdashtype{gp dt solid}
\gpsetlinewidth{1.00}
\draw[gp path] (1.320,2.575)--(1.500,2.575);
\node[gp node right] at (1.136,2.575) {$0.6$};
\gpcolor{color=gp lt color axes}
\gpsetlinetype{gp lt axes}
\gpsetdashtype{gp dt axes}
\gpsetlinewidth{0.50}
\draw[gp path] (1.320,3.105)--(7.947,3.105);
\gpcolor{color=gp lt color border}
\gpsetlinetype{gp lt border}
\gpsetdashtype{gp dt solid}
\gpsetlinewidth{1.00}
\draw[gp path] (1.320,3.105)--(1.500,3.105);
\node[gp node right] at (1.136,3.105) {$0.8$};
\gpcolor{color=gp lt color axes}
\gpsetlinetype{gp lt axes}
\gpsetdashtype{gp dt axes}
\gpsetlinewidth{0.50}
\draw[gp path] (1.320,3.636)--(7.947,3.636);
\gpcolor{color=gp lt color border}
\gpsetlinetype{gp lt border}
\gpsetdashtype{gp dt solid}
\gpsetlinewidth{1.00}
\draw[gp path] (1.320,3.636)--(1.500,3.636);
\node[gp node right] at (1.136,3.636) {$1$};
\gpcolor{color=gp lt color axes}
\gpsetlinetype{gp lt axes}
\gpsetdashtype{gp dt axes}
\gpsetlinewidth{0.50}
\draw[gp path] (1.320,4.166)--(7.947,4.166);
\gpcolor{color=gp lt color border}
\gpsetlinetype{gp lt border}
\gpsetdashtype{gp dt solid}
\gpsetlinewidth{1.00}
\draw[gp path] (1.320,4.166)--(1.500,4.166);
\node[gp node right] at (1.136,4.166) {$1.2$};
\gpcolor{color=gp lt color axes}
\gpsetlinetype{gp lt axes}
\gpsetdashtype{gp dt axes}
\gpsetlinewidth{0.50}
\draw[gp path] (1.320,4.696)--(5.743,4.696);
\draw[gp path] (7.763,4.696)--(7.947,4.696);
\gpcolor{color=gp lt color border}
\gpsetlinetype{gp lt border}
\gpsetdashtype{gp dt solid}
\gpsetlinewidth{1.00}
\draw[gp path] (1.320,4.696)--(1.500,4.696);
\node[gp node right] at (1.136,4.696) {$1.4$};
\gpcolor{color=gp lt color axes}
\gpsetlinetype{gp lt axes}
\gpsetdashtype{gp dt axes}
\gpsetlinewidth{0.50}
\draw[gp path] (1.320,5.226)--(5.743,5.226);
\draw[gp path] (7.763,5.226)--(7.947,5.226);
\gpcolor{color=gp lt color border}
\gpsetlinetype{gp lt border}
\gpsetdashtype{gp dt solid}
\gpsetlinewidth{1.00}
\draw[gp path] (1.320,5.226)--(1.500,5.226);
\node[gp node right] at (1.136,5.226) {$1.6$};
\gpcolor{color=gp lt color axes}
\gpsetlinetype{gp lt axes}
\gpsetdashtype{gp dt axes}
\gpsetlinewidth{0.50}
\draw[gp path] (1.320,5.756)--(7.947,5.756);
\gpcolor{color=gp lt color border}
\gpsetlinetype{gp lt border}
\gpsetdashtype{gp dt solid}
\gpsetlinewidth{1.00}
\draw[gp path] (1.320,5.756)--(1.500,5.756);
\node[gp node right] at (1.136,5.756) {$1.8$};
\draw[gp path] (1.320,0.985)--(1.320,1.165);
\node[gp node center] at (1.320,0.677) {$16$};
\draw[gp path] (2.977,0.985)--(2.977,1.165);
\node[gp node center] at (2.977,0.677) {$32$};
\draw[gp path] (4.634,0.985)--(4.634,1.165);
\node[gp node center] at (4.634,0.677) {$64$};
\draw[gp path] (6.290,0.985)--(6.290,1.165);
\node[gp node center] at (6.290,0.677) {$128$};
\draw[gp path] (7.947,0.985)--(7.947,1.165);
\node[gp node center] at (7.947,0.677) {$256$};
\draw[gp path] (1.320,5.756)--(1.320,0.985)--(7.947,0.985)--(7.947,5.756)--cycle;
\node[gp node center,rotate=-270] at (0.246,3.370) {Time\textsubscript{~DAOS-Chunked} / Time\textsubscript{~Lustre}};
\node[gp node center] at (4.633,0.215) {Number of Processes};
\node[gp node center] at (6.753,5.422) {Problem Size};
\node[gp node right] at (6.479,5.114) {512};
\gpcolor{rgb color={0.216,0.494,0.722}}
\gpsetlinewidth{1.80}
\draw[gp path] (6.663,5.114)--(7.579,5.114);
\draw[gp path] (1.320,3.365)--(2.977,2.490)--(4.634,1.600)--(6.290,1.444)--(7.947,1.439);
\gpsetpointsize{7.20}
\gppoint{gp mark 6}{(1.320,3.365)}
\gppoint{gp mark 6}{(2.977,2.490)}
\gppoint{gp mark 6}{(4.634,1.600)}
\gppoint{gp mark 6}{(6.290,1.444)}
\gppoint{gp mark 6}{(7.947,1.439)}
\gppoint{gp mark 6}{(7.121,5.114)}
\draw[gp path] (1.320,3.356)--(1.320,3.531);
\draw[gp path] (1.320,3.356)--(1.410,3.356);
\draw[gp path] (1.320,3.531)--(1.410,3.531);
\draw[gp path] (2.977,2.488)--(2.977,2.553);
\draw[gp path] (2.887,2.488)--(3.067,2.488);
\draw[gp path] (2.887,2.553)--(3.067,2.553);
\draw[gp path] (4.634,1.599)--(4.634,1.625);
\draw[gp path] (4.544,1.599)--(4.724,1.599);
\draw[gp path] (4.544,1.625)--(4.724,1.625);
\draw[gp path] (6.290,1.443)--(6.290,1.458);
\draw[gp path] (6.200,1.443)--(6.380,1.443);
\draw[gp path] (6.200,1.458)--(6.380,1.458);
\draw[gp path] (7.947,1.438)--(7.947,1.448);
\draw[gp path] (7.857,1.438)--(7.947,1.438);
\draw[gp path] (7.857,1.448)--(7.947,1.448);
\gppoint{gp mark 6}{(1.320,3.365)}
\gppoint{gp mark 6}{(2.977,2.490)}
\gppoint{gp mark 6}{(4.634,1.600)}
\gppoint{gp mark 6}{(6.290,1.444)}
\gppoint{gp mark 6}{(7.947,1.439)}
\gpcolor{color=gp lt color border}
\node[gp node right] at (6.479,4.806) {1024};
\gpcolor{rgb color={0.894,0.102,0.110}}
\draw[gp path] (6.663,4.806)--(7.579,4.806);
\draw[gp path] (1.320,4.469)--(2.977,4.423)--(4.634,1.819)--(6.290,1.541)--(7.947,1.428);
\gppoint{gp mark 4}{(1.320,4.469)}
\gppoint{gp mark 4}{(2.977,4.423)}
\gppoint{gp mark 4}{(4.634,1.819)}
\gppoint{gp mark 4}{(6.290,1.541)}
\gppoint{gp mark 4}{(7.947,1.428)}
\gppoint{gp mark 4}{(7.121,4.806)}
\draw[gp path] (1.320,4.460)--(1.320,4.650);
\draw[gp path] (1.320,4.460)--(1.410,4.460);
\draw[gp path] (1.320,4.650)--(1.410,4.650);
\draw[gp path] (2.977,4.418)--(2.977,4.544);
\draw[gp path] (2.887,4.418)--(3.067,4.418);
\draw[gp path] (2.887,4.544)--(3.067,4.544);
\draw[gp path] (4.634,1.818)--(4.634,1.848);
\draw[gp path] (4.544,1.818)--(4.724,1.818);
\draw[gp path] (4.544,1.848)--(4.724,1.848);
\draw[gp path] (6.290,1.540)--(6.290,1.562);
\draw[gp path] (6.200,1.540)--(6.380,1.540);
\draw[gp path] (6.200,1.562)--(6.380,1.562);
\draw[gp path] (7.947,1.427)--(7.947,1.436);
\draw[gp path] (7.857,1.427)--(7.947,1.427);
\draw[gp path] (7.857,1.436)--(7.947,1.436);
\gppoint{gp mark 4}{(1.320,4.469)}
\gppoint{gp mark 4}{(2.977,4.423)}
\gppoint{gp mark 4}{(4.634,1.819)}
\gppoint{gp mark 4}{(6.290,1.541)}
\gppoint{gp mark 4}{(7.947,1.428)}
\gpcolor{color=gp lt color border}
\node[gp node right] at (6.479,4.498) {2048};
\gpcolor{rgb color={0.400,0.400,0.400}}
\draw[gp path] (6.663,4.498)--(7.579,4.498);
\draw[gp path] (1.320,5.263)--(2.977,4.675)--(4.634,1.911)--(6.290,1.593)--(7.947,1.620);
\gppoint{gp mark 8}{(1.320,5.263)}
\gppoint{gp mark 8}{(2.977,4.675)}
\gppoint{gp mark 8}{(4.634,1.911)}
\gppoint{gp mark 8}{(6.290,1.593)}
\gppoint{gp mark 8}{(7.947,1.620)}
\gppoint{gp mark 8}{(7.121,4.498)}
\draw[gp path] (1.320,5.257)--(1.320,5.368);
\draw[gp path] (1.320,5.257)--(1.410,5.257);
\draw[gp path] (1.320,5.368)--(1.410,5.368);
\draw[gp path] (2.977,4.673)--(2.977,4.763);
\draw[gp path] (2.887,4.673)--(3.067,4.673);
\draw[gp path] (2.887,4.763)--(3.067,4.763);
\draw[gp path] (4.634,1.911)--(4.634,1.936);
\draw[gp path] (4.544,1.911)--(4.724,1.911);
\draw[gp path] (4.544,1.936)--(4.724,1.936);
\draw[gp path] (6.290,1.592)--(6.290,1.608);
\draw[gp path] (6.200,1.592)--(6.380,1.592);
\draw[gp path] (6.200,1.608)--(6.380,1.608);
\draw[gp path] (7.947,1.620)--(7.947,1.628);
\draw[gp path] (7.857,1.620)--(7.947,1.620);
\draw[gp path] (7.857,1.628)--(7.947,1.628);
\gppoint{gp mark 8}{(1.320,5.263)}
\gppoint{gp mark 8}{(2.977,4.675)}
\gppoint{gp mark 8}{(4.634,1.911)}
\gppoint{gp mark 8}{(6.290,1.593)}
\gppoint{gp mark 8}{(7.947,1.620)}
\gpcolor{color=gp lt color border}
\gpsetlinewidth{1.00}
\draw[gp path] (1.320,5.756)--(1.320,0.985)--(7.947,0.985)--(7.947,5.756)--cycle;
%% coordinates of the plot area
\gpdefrectangularnode{gp plot 1}{\pgfpoint{1.320cm}{0.985cm}}{\pgfpoint{7.947cm}{5.756cm}}
\end{tikzpicture}
%% gnuplot variables
\par\end{centering}
\vspace{-15pt}
\caption{\label{fig:CLAMR_daoschunked_lustre}
Comparison of chunked CLAMR I/O on DAOS to unchunked CLAMR I/O on Lustre.  Values above one represent chunked CLAMR I/O on DAOS performing worse than unchunked CLAMR I/O on Lustre.}
\end{figure}

CLAMR write performance was then assessed using the DAOS stack both with and without
the use of chunking. Using the previous data gathered for CLAMR's performance
in writing checkpoint files to a Lustre filesystem, a comparison was drawn,
showing that DAOS without chunking performs twice as well on average as
compared to Lustre (up to nearly three times as well in specific cases) when
the number of processors is sufficiently large, Fig. \ref{fig:CLAMR_daos_lustre}.
In order to assess the performance impact that chunking has, $8$ DAOS servers
were started across $8$ different nodes and a variety of chunk sizes ranging
from $2^{9}$ to $2^{21}$ were tested. When chunking is enabled, DAOS performance
can exceed the unchunked performance by approximately $25$--$50$\%,
depending on the number of processors used and the appropriateness of the chunk
size chosen for the given problem size, Fig. \ref{fig:CLAMR_daos_daoschunked}.
When the number of processors used is large enough, this in turn leads to a
consistent $50$--$75$\% increase in performance as compared to Lustre
unchunked I/O, Fig. \ref{fig:CLAMR_daoschunked_lustre}.

\subsection{\label{subsec:Legion-parallel-programming}Legion Parallel Programming
System}

Legion~\cite{Bauer2012} is a high-level data-centric and task-based
application runtime. Legion's primary data model is based on \textsl{Logical
Regions} which are the cross product of an \textsl{N}-dimensional
index space and a multi-variable field space. Logical regions are
distinct from the physical regions (memories) that underlie the logical
region and provide the physical instantiation of the data. The data
model also provides a method of coloring the index space and/or the
field space, which can be used to partition an index space or slice
the field space of the logical region. The runtime can then use this
coloring and a partitioning applied to the logical region to manage
a distributed instance of the logical region.

Legion is currently capable of attaching to HDF5 files using the low-level
Realm runtime on which Legion is built. Realm maps HDF5 files (or
groups) to memory objects within the runtime using a Direct Memory Access (DMA) mechanism
to collect updates to individual datasets in an HDF5 file. As part
of demonstrating Legion's use of HDF5 on the DAOS storage stack, an
example application is created that demonstrates the capabilities
of the storage stack. \texttt{Tester\_io}\footnote{\texttt{Tester\_io} is part
of Legion's Github repository and can be found in the source tree
at test/hdf\_attach\_subregion\_parallel} is a straightforward
tour of the HDF5-DAOS features from a simple Legion code and is designed
to run the Legion runtime on multiple compute nodes, talking to multiple
I/O and storage server nodes.

In the current implementation of HDF5 on top of DAOS, transactions
and the event stack are handled internally in the HDF5 library, and
consequently, no extra parameters needed to be added to the original HDF5 APIs. Therefore, the number of changes from the original Legion
and HDF5 implementation was minimal. One new HDF5 API call, \texttt{H5VLdaosm\_init}, was introduced
into Legion, which initializes the VOL plugin by connecting to the DAOS server pool and registering the driver with the HDF5 library. This function is called only once by all the processes within Legion.

The Legion benchmark relies on the low-level networking layer \texttt{gasnet}~\cite{Bonachea2002}
for network-independent, high-performance communication
primitives. The \texttt{mpiexec} command in the gasnet wrapper script
\texttt{gasnetrun\_ibv} needed to be updated to include DAOS specific parameters.
The command for running \texttt{tester\_io} on $3$ nodes with $131072$ elements
and $256$ shards is:
\begin{tcolorbox}[colback=white,left=5pt,opacityframe=0.5]
\scriptsize
\begin{verbatim}
$ GASNET_BACKTRACE=1 GASNET_USE_XRC=0 \
    GASNET_MASTERIP='...' GASNET_SPAWN=-L \
    gasnetrun_ibv -n 3 tester_io -n 131072 -s 256
\end{verbatim}
\end{tcolorbox}
As was the case with CLAMR, the \textit{pool id} is passed to Legion
through the environment variable \texttt{pid}.

\begin{figure}
\begin{centering}
\includegraphics[width=1\linewidth]{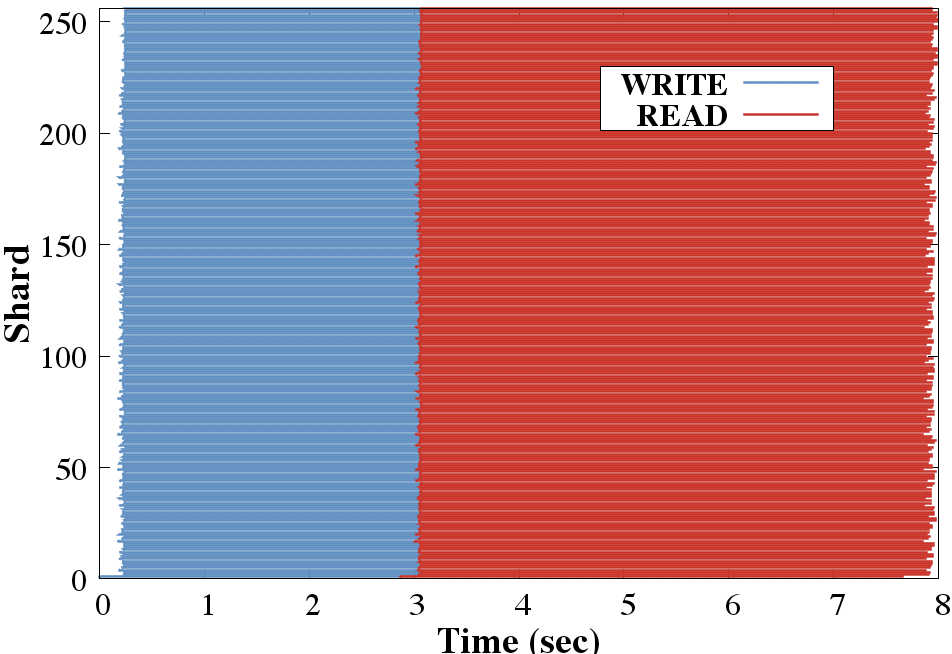} 
\par\end{centering}
%\vspace{5pt}
\caption{\label{fig:Time-series-of-the}Time-series of the I/O phases (write
and read) associated with each shard's global data structure being
persisted for HDF5 with DAOS, highlighting Legion's ability to perform
independent I/O phases associated with each shard of global data for
DAOS.}
\end{figure}

The non-bulk synchronous ability of DAOS via the transaction and epoch
model of DAOS was demonstrated bu using \texttt{Tester\_io}. Fig. \ref{fig:Time-series-of-the}
shows the typical workload for the read/write phase for different
shards in \texttt{Tester\_io}. Additionally, Fig. \ref{fig:Time-series-of-the}
highlights Legion's capability of scheduling different phases of tasks
based on explicit dependencies and, consequently, allows for reading
tasks to run concurrently with writing tasks on the same logical region~\cite{Watkins2015}.

\begin{figure}
\begin{centering}
\input{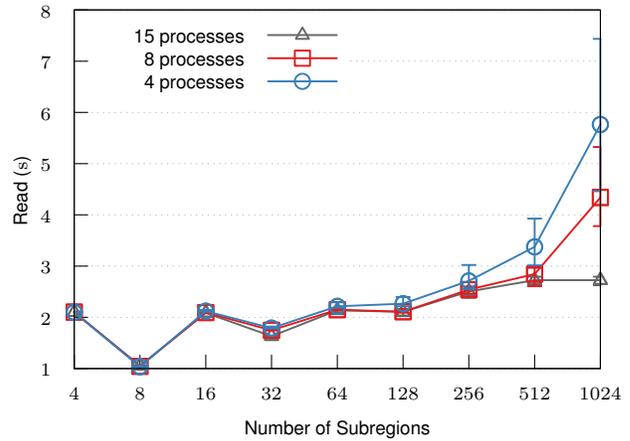}
\par\end{centering}
\vspace{-15pt}
\caption{\label{fig:DAOS-read-performances}DAOS read performances as the number
of MPI processes is increased from 4 to 15 processes for a different
number of subregions.}
\end{figure}

\texttt{Tester\_io} was used to study the effects of the number of MPI processes
as a function of the number of Legion subregions, where the size of
the problem was fixed for the number of elements of $2^{29}$. The number of DAOS servers was one and the
number of Legion MPI processes was 4, 8 and 15 where each compute node executed
only one MPI process, and each node could run 74 tasks. The higher number of process made a large 
difference when the number
of subregions increased to greater than 256 for both reading and writing,
Fig. \ref{fig:DAOS-read-performances} and Fig. \ref{fig:DAOS-write-performances}.

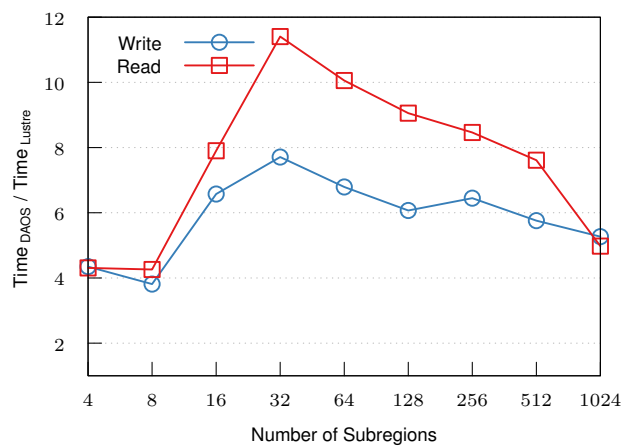
\begin{figure}
\begin{centering}
\begin{tikzpicture}[gnuplot]
%% generated with GNUPLOT 5.0p5 (Lua 5.3; terminal rev. 99, script rev. 100)
%% Thu 29 Jun 2017 01:59:32 PM CDT
\tikzset{every node/.append style={font={\sffamily \scriptsize}}}
\path (0.000,0.000) rectangle (8.500,6.125);
\gpcolor{color=gp lt color axes}
\gpsetlinetype{gp lt axes}
\gpsetdashtype{gp dt axes}
\gpsetlinewidth{0.50}
\draw[gp path] (1.136,1.419)--(7.947,1.419);
\gpcolor{color=gp lt color border}
\gpsetlinetype{gp lt border}
\gpsetdashtype{gp dt solid}
\gpsetlinewidth{1.00}
\draw[gp path] (1.136,1.419)--(1.316,1.419);
\node[gp node right] at (0.952,1.419) {$2$};
\gpcolor{color=gp lt color axes}
\gpsetlinetype{gp lt axes}
\gpsetdashtype{gp dt axes}
\gpsetlinewidth{0.50}
\draw[gp path] (1.136,2.286)--(7.947,2.286);
\gpcolor{color=gp lt color border}
\gpsetlinetype{gp lt border}
\gpsetdashtype{gp dt solid}
\gpsetlinewidth{1.00}
\draw[gp path] (1.136,2.286)--(1.316,2.286);
\node[gp node right] at (0.952,2.286) {$4$};
\gpcolor{color=gp lt color axes}
\gpsetlinetype{gp lt axes}
\gpsetdashtype{gp dt axes}
\gpsetlinewidth{0.50}
\draw[gp path] (1.136,3.154)--(7.947,3.154);
\gpcolor{color=gp lt color border}
\gpsetlinetype{gp lt border}
\gpsetdashtype{gp dt solid}
\gpsetlinewidth{1.00}
\draw[gp path] (1.136,3.154)--(1.316,3.154);
\node[gp node right] at (0.952,3.154) {$6$};
\gpcolor{color=gp lt color axes}
\gpsetlinetype{gp lt axes}
\gpsetdashtype{gp dt axes}
\gpsetlinewidth{0.50}
\draw[gp path] (1.136,4.021)--(7.947,4.021);
\gpcolor{color=gp lt color border}
\gpsetlinetype{gp lt border}
\gpsetdashtype{gp dt solid}
\gpsetlinewidth{1.00}
\draw[gp path] (1.136,4.021)--(1.316,4.021);
\node[gp node right] at (0.952,4.021) {$8$};
\gpcolor{color=gp lt color axes}
\gpsetlinetype{gp lt axes}
\gpsetdashtype{gp dt axes}
\gpsetlinewidth{0.50}
\draw[gp path] (1.136,4.889)--(7.947,4.889);
\gpcolor{color=gp lt color border}
\gpsetlinetype{gp lt border}
\gpsetdashtype{gp dt solid}
\gpsetlinewidth{1.00}
\draw[gp path] (1.136,4.889)--(1.316,4.889);
\node[gp node right] at (0.952,4.889) {$10$};
\gpcolor{color=gp lt color axes}
\gpsetlinetype{gp lt axes}
\gpsetdashtype{gp dt axes}
\gpsetlinewidth{0.50}
\draw[gp path] (1.136,5.756)--(7.947,5.756);
\gpcolor{color=gp lt color border}
\gpsetlinetype{gp lt border}
\gpsetdashtype{gp dt solid}
\gpsetlinewidth{1.00}
\draw[gp path] (1.136,5.756)--(1.316,5.756);
\node[gp node right] at (0.952,5.756) {$12$};
\draw[gp path] (1.136,0.985)--(1.136,1.165);
\node[gp node center] at (1.136,0.677) {$4$};
\draw[gp path] (1.987,0.985)--(1.987,1.165);
\node[gp node center] at (1.987,0.677) {$8$};
\draw[gp path] (2.839,0.985)--(2.839,1.165);
\node[gp node center] at (2.839,0.677) {$16$};
\draw[gp path] (3.690,0.985)--(3.690,1.165);
\node[gp node center] at (3.690,0.677) {$32$};
\draw[gp path] (4.542,0.985)--(4.542,1.165);
\node[gp node center] at (4.542,0.677) {$64$};
\draw[gp path] (5.393,0.985)--(5.393,1.165);
\node[gp node center] at (5.393,0.677) {$128$};
\draw[gp path] (6.244,0.985)--(6.244,1.165);
\node[gp node center] at (6.244,0.677) {$256$};
\draw[gp path] (7.096,0.985)--(7.096,1.165);
\node[gp node center] at (7.096,0.677) {$512$};
\draw[gp path] (7.947,0.985)--(7.947,1.165);
\node[gp node center] at (7.947,0.677) {$1024$};
\draw[gp path] (1.136,5.756)--(1.136,0.985)--(7.947,0.985)--(7.947,5.756)--cycle;
\node[gp node center,rotate=-270] at (0.246,3.370) {Time\textsubscript{~DAOS} / Time\textsubscript{~Lustre}};
\node[gp node center] at (4.541,0.215) {Number of Subregions};
\node[gp node right] at (2.240,5.422) {Write};
\gpcolor{rgb color={0.216,0.494,0.722}}
\gpsetlinewidth{1.80}
\draw[gp path] (2.424,5.422)--(3.340,5.422);
\draw[gp path] (1.136,2.437)--(1.987,2.205)--(2.839,3.403)--(3.690,3.895)--(4.542,3.497)%
  --(5.393,3.184)--(6.244,3.348)--(7.096,3.049)--(7.947,2.836);
\gpsetpointsize{7.20}
\gppoint{gp mark 6}{(1.136,2.437)}
\gppoint{gp mark 6}{(1.987,2.205)}
\gppoint{gp mark 6}{(2.839,3.403)}
\gppoint{gp mark 6}{(3.690,3.895)}
\gppoint{gp mark 6}{(4.542,3.497)}
\gppoint{gp mark 6}{(5.393,3.184)}
\gppoint{gp mark 6}{(6.244,3.348)}
\gppoint{gp mark 6}{(7.096,3.049)}
\gppoint{gp mark 6}{(7.947,2.836)}
\gppoint{gp mark 6}{(2.882,5.422)}
\gpcolor{color=gp lt color border}
\node[gp node right] at (2.240,5.114) {Read};
\gpcolor{rgb color={0.894,0.102,0.110}}
\draw[gp path] (2.424,5.114)--(3.340,5.114);
\draw[gp path] (1.136,2.420)--(1.987,2.399)--(2.839,3.979)--(3.690,5.498)--(4.542,4.913)%
  --(5.393,4.479)--(6.244,4.222)--(7.096,3.853)--(7.947,2.710);
\gppoint{gp mark 4}{(1.136,2.420)}
\gppoint{gp mark 4}{(1.987,2.399)}
\gppoint{gp mark 4}{(2.839,3.979)}
\gppoint{gp mark 4}{(3.690,5.498)}
\gppoint{gp mark 4}{(4.542,4.913)}
\gppoint{gp mark 4}{(5.393,4.479)}
\gppoint{gp mark 4}{(6.244,4.222)}
\gppoint{gp mark 4}{(7.096,3.853)}
\gppoint{gp mark 4}{(7.947,2.710)}
\gppoint{gp mark 4}{(2.882,5.114)}
\gpcolor{color=gp lt color border}
\gpsetlinewidth{1.00}
\draw[gp path] (1.136,5.756)--(1.136,0.985)--(7.947,0.985)--(7.947,5.756)--cycle;
%% coordinates of the plot area
\gpdefrectangularnode{gp plot 1}{\pgfpoint{1.136cm}{0.985cm}}{\pgfpoint{7.947cm}{5.756cm}}
\end{tikzpicture}
%% gnuplot variables 
\par\end{centering}
\vspace{-15pt}
\caption{\label{fig:DAOS-read-and}DAOS read and write performance in comparison
to Lustre. DAOS uses a single service thread while Lustre uses multiple threads on multiple servers.}
\end{figure}

The performance of Legion using DAOS and Lustre is presented in Fig. \ref{fig:DAOS-read-and}
for 15 MPI processes and where the number of elements is $2^{30}$. The slowdown in DAOS and the big performance hit when compared with Lustre in this case was expected, since the Legion tester code generates a large number of I/O calls, and the HDF5 implementation in Legion did not use chunking layout for the Datasets. This results in all the I/O calls being serialized at the DAOS server to one service thread, whereas the Lustre server utilizes many threads to handle the incoming I/Os. Furthermore, as mentioned earlier, the network interface used by DAOS on this cluster was libfabric over TCP which is slower compared to InfiniBand verbs utilized by Lustre. We also noticed that the \textit{tmpfs} file system was returning out of space errors for larger problem sizes, even when space was available, after a certain number of memory allocations triggered by the large number of I/O operations from the client. All those issues will be addressed in future development of both DAOS and the HDF5 DAOS backend.

\begin{figure}
\begin{centering}
\input{LegionWrite}
\par\end{centering}
\vspace{-15pt}
\caption{\label{fig:DAOS-write-performances}DAOS write performances as the
number of MPI processes is increased from 4 to 15 processes for a
different number of subregions.}
\end{figure}

\subsection{High-level I/O Stack Libraries\label{subsec:HL}}

There is a desire to support legacy libraries on the DAOS stack that
already make use of HDF5. The objective is to have mid-level
code completely manage DAOS, isolating the application code from
the I/O stack.

\subsubsection{\label{subsec:DAOSNETCDF}DAOS NetCDF Implementation}

NetCDF~\cite{NetCDF2016} is a set of software libraries used to facilitate
the creation, access, and sharing of array-oriented scientific data
in self-describing, machine-independent data formats. A new set of
DAOS NetCDF APIs were derived from the original NetCDF APIs, maintaining
most of the original functionality. 

The non-DAOS version of NetCDF added dimensions to variables by the
use of HDF5 \textit{Dimension Scale} APIs. Storing dimensions with
coordinate variables (variables used as a dimension scale for a dimension
of the same name) is intuitive and is self-describing for applications
that access the file directly through HDF5. However, this approach
introduces several dependencies between datasets and attributes that
do not fit well with the DAOS transaction model. In addition, it introduces
many special cases that must be handled, increasing the difficulty
of implementation. Finally, there is currently no DAOS implementation
of the Dimension Scale APIs and implementing them would be difficult
due to the transaction model.

\begin{figure}
\begin{centering}
\includegraphics[width=1\linewidth]{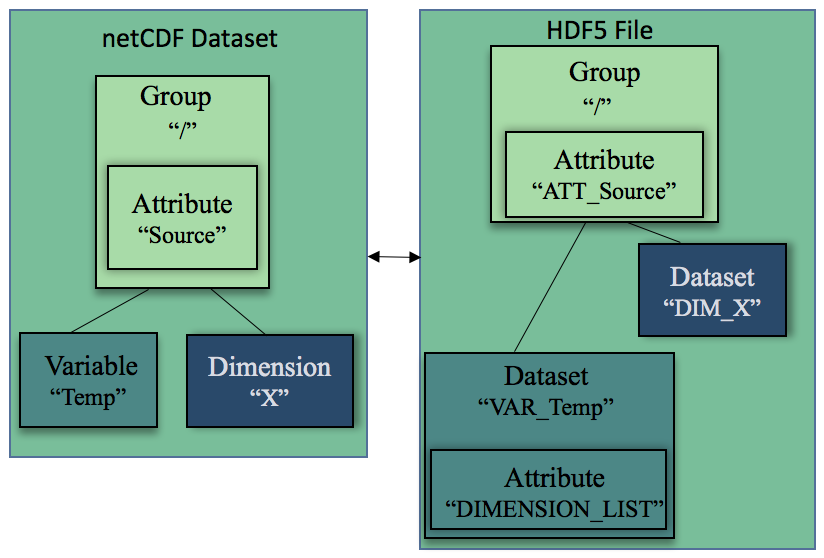} 
\par\end{centering}
\caption{\label{fig:HDF5-schema-for}HDF5 schema for NetCDF/DAOS.}
\end{figure}

Since backward compatibility with NetCDF's file format was not of a concern (i.e., having NetCDF/DAOS
datasets/containers/files being accessed independently of the NetCDF/DAOS
API) the existing NetCDF4 schema for HDF5 was abandoned in order to
simplify the implementation. All variables and dimensions were implemented
as HDF5 datasets, all groups as HDF5 groups, and all attributes as
HDF5 attributes. As a convention, all dimensions have the string DIM\_
prepended to the name in HDF5, all variables have the string VAR\_
prepended, and all attributes have the string ATT\_ prepended. Variables
have an HDF5 attribute DIMENSION\_LIST, invisible to the NetCDF API,
that stores references to the dimensions for the variable, Fig. \ref{fig:HDF5-schema-for}.
Dimensions are implemented as a scalar dataset of type H5T\_STD\_U64LE,
where the value indicates the dimension length or all \textsc{1}s
(i.e. \texttt{(uint64\_t)(int64\_t)-1}) to indicate an unlimited dimension.
This implementation avoids all name conflicts without having to add
any special cases to the code, and also allows the removal of code
paths for handling coordinate variables as a special case, instead
of treating them like any other variable. All the NetCDF APIs that use this new schema were appended with a ``\texttt{\_ff}'' in their names. Other DAOS additions to
NetCDF included:
\begin{itemize}
\item Support for unlimited dimensions, but only for collective access and
only for the slowest changing dimension;
\item ``Links'' from variables to their dimensions, allowing the variables
to be queried about their dimensions.
\end{itemize}

\subsubsection{DAOS\emph{ Parallel I/O} (PIO) Implementation}

A common application library which uses NetCDF is the software associated
with the \textbf{A}ccelerated \textbf{C}limate \textbf{M}odeling for
\textbf{E}nergy (ACME)~\cite{ACME}program. ACME uses the package
\emph{Parallel I/O} (PIO)~\cite{PIO} to perform I/O which, in turn,
uses as its backend the NetCDF file format. Since the global DAOS
stack variables are isolated from both PIO and NetCDF, PIO APIs simply
use the DAOS ``\texttt{\_ff}'' NetCDF APIs mentioned in Section \ref{subsec:DAOSNETCDF}. The transaction number is automatically initialized and incremented as needed within HDF5. Similar to the NetCDF convention, all new DAOS PIO C APIs are indicated by appending a ``\texttt{\_ff}'' to the function
names.

PIO expects as input from the application the partitioned data arrays
for each process. Additionally, PIO has the option for requesting
a subset of the CN that will perform the I/O. Hence, PIO aggregates
the I/O from each process to only a subset of processes for I/O. The
I/O processes then use NetCDF APIs to carry out the I/O. PIO implements
two methods for aggregating the I/O from all the processes to the
subset of I/O processes. In the box method, each compute task will
transfer data to one or more of the I/O processes. For the subset
method, each I/O process is associated with a unique subset of compute
processes for which each compute process transfers data to only one
I/O process~\cite{Edwards2016}. In general, the subset method reduces
the overall communication cost when compared to the box method.

Additionally, since PIO has the capability of using a subset of processes
for I/O, \texttt{H5VLdaosm\_init} (i.e., a DAOS HDF5 API used to start
the DAOS stack) uses the MPI sub-communicator group so that only those
processes involved in I/O will initialize the DAOS stack. This initialization
of the DAOS stack happens automatically when the I/O MPI sub-communicator
is created in PIO and it is finalized when this same sub-communicator
is freed in PIO.

PIO's testing program \texttt{pioperformance.F90} uses two input files;
the first file contains namelist settings for the testing parameters
and the second file contains the decomposition information from a
PIO program (e.g. CESM, ACME). The test program reads namelist and
then generates test data consisting of integers, 4-byte reals and
8-byte reals. It then writes the data using DAOS NetCDF via PIO APIs
and then reads the data back using DAOS PIO, checks for correctness,
and outputs the data rate in reading and writing the data.

\begin{figure}
\begin{centering}
\includegraphics[width=1\linewidth]{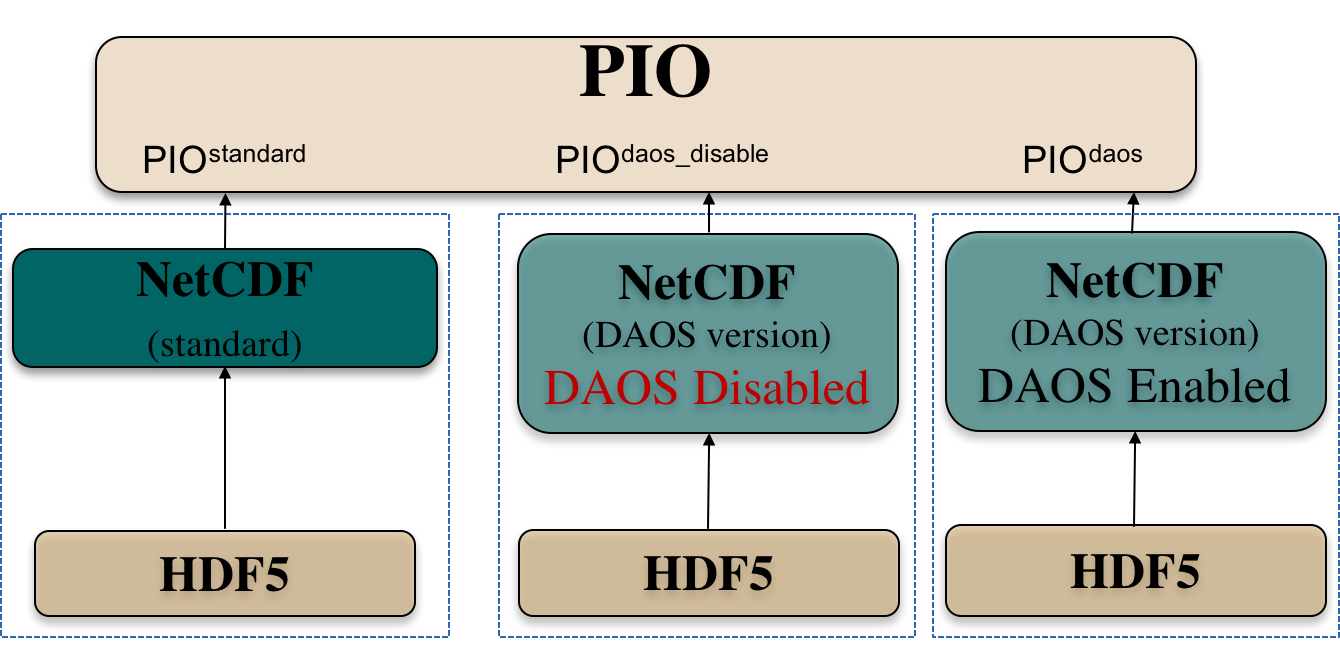} 
\par\end{centering}
\vspace{-5pt}
\caption{\label{fig:Three-combinations-of}Three combinations of PIO, NetCDF
and HDF5.}
\end{figure}

There are two versions of NetCDF: (1) the ``standard'' version is
the unmodified v4.4.1 of NetCDF available from Unidata and (2) the
``DAOS'' version, which is a modified v4.4.1 of NetCDF, as previously
discussed. Three combinations of PIO, NetCDF and HDF5 are investigated,
Fig. \ref{fig:Three-combinations-of}:
\begin{enumerate}
\item PIO\textsuperscript{standard}---The PIO configuration
uses standard HDF5 with the standard version of NetCDF;
\item PIO\textsuperscript{daos\_disable}---The PIO configuration
uses standard HDF5 with the DAOS version of NetCDF, where the DAOS
capabilities in NetCDF are disabled;
\item PIO\textsuperscript{daos}---The PIO configuration uses
the DAOS version of HDF5, and the DAOS capabilities in NetCDF are
enabled. 
\end{enumerate}
Thus, for both case 1 and 2, a POSIX file is getting written, via
HDF5, to a Lustre backend.

All tests were made on Boro to compare PIO with Lustre and DAOS, and
to verify the DAOS PIO implementation. An investigation of the Lustre
parameters resulted in a stripe count of 1 with a stripe size of 2MB
being used for all the Lustre benchmark results. The symbol in the
figures is obtained by running the benchmarks ten times and averaging
the I/O times over those ten runs. The vertical line segment represents
the minimum and maximum of I/O over the ten runs.

%\begin{minipage}[c][1\totalheight][t]{0.45\textwidth}%
\begin{figure}
\begin{centering}
\input{PIORead}
\par\end{centering}
\vspace{-15pt}
\caption{\label{fig:Read-performance-of}Read performance comparison between Lustre and DAOS for 1024 processes. }
%\end{minipage}

%\begin{minipage}[c][1\totalheight][t]{0.45\textwidth}%
\begin{centering}
\input{PIOWrite}
\par\end{centering}
\vspace{-15pt}
\caption{\label{fig:Write-erformance-of}Nearly identical DAOS and Lustre write performance for 1024 processes.}
%\end{minipage}
\end{figure}

The benchmark used 1024 processes and the decomposition file, \texttt{piodecomp1024tasks03dims05.dat}.
The modified DAOS version of NetCDF outperforms the standard NetCDF version
for both reading (Fig. \ref{fig:Read-performance-of}), and writing (Fig. \ref{fig:Write-erformance-of}). Moreover, the DOAS implementation is on average faster than the Lustre implementation for reading and matches Lustre complete times for writing.

\section{Conclusions}
The increase from petascale to exascale for storage and I/O requires
a new approach to the architecture because it is not possible to just
scale prior systems. As shown, DAOS is primarily designed for these next generation systems which use  NVRAM and NVMe storage technology as a means to provide a high bandwidth storage tier very close to the compute node.

This research highlights the challenges and issues when porting existing applications codes to a transaction model for I/O. The transition to using DAOS for application and middleware developers is made easier by using HDF5, which hides the DAOS schema within HDF5. Although this paper focuses on HDF5 and DAOS, other I/O libraries such as MPI-I/O and POSIX I/O will also be supported on top of DAOS in the future.

The performance of DAOS in comparison to Lustre from a diverse set of applications was highlighted. Although the comparison was not justified due to the significant differences in the hardware and software stack, in similar workloads DAOS completed either faster or equivalent to Lustre. Other cases highlight the importance of leveraging multiple DAOS server processes and threads to avoid the bottleneck at the server side when performing large number of I/Os from the client without any aggregation, and provided ground for improving both the DAOS implementation and the HDF5 plugin for DAOS, as both are still in the prototyping phase.


\ifCLASSOPTIONcompsoc
  \section*{Acknowledgments}
\else
  \section*{Acknowledgment}
\fi

Funding for this project was provided through Intel subcontract \#CW1998599
from Intel's contract \#B613306 with Lawrence Livermore National Security. Access to the Boro cluster was provided by Intel.



\bibliographystyle{IEEEtran}
\bibliography{IEEEabrv,FFapps}
%\begin{thebibliography}{1}

%\bibitem[Edwards et~al., 2016]{Edwards2016}
%Edwards, J., Dennis, J.~M., and Vertenstein, M. (2016).
%\newblock {Parallel I/O library (PIO)}.
%\newblock {\em http://ncar.github.io/ParallelIO/}.

%\bibitem[Habib et~al., 2014]{Habib2014}
%Habib, S., Pope, A., Finkel, H., Frontiere, N., Heitmann, K., Daniel, D.,
%  Fasel, P., Morozov, V., Zagaris, G., Peterka, T., Vishwanath, V., Lukic, Z.,
%  Sehrish, S., and Liao, W.-k. (2014).
%\newblock {HACC: Simulating Sky Surveys on State-of-the-Art Supercomputing
%  Architectures}.
%\newblock 4.

%\bibitem[Isaila et~al., 2016]{Isaila2016}
%Isaila, F., Garcia, J., Carretero, J., Ross, R., and Kimpe, D. (2016).
%\newblock {Making the case for reforming the I/O software stack of
%  extreme-scale systems}.
%\newblock {\em Advances in Engineering Software}, 16(10):1--6.

%\bibitem[Keyes, 2011]{Keyes2011}
%Keyes, D.~E. (2011).
%\newblock {Exaflop/s: The why and the how}.
%\newblock {\em Comptes Rendus M{\'{e}}canique}, 339(2-3):70--77.

%\bibitem[Kogge et~al., 2008]{Kogge2008}
%Kogge, P., Bergman, K., Borkar, S., Campbell, D., Carson, W., Dally, W.,
%  Denneau, M., Franzon, P., Harrod, W., Hill, K., Hiller, J., Richards, M., and
%  Snavely, A. (2008).
%\newblock {ExaScale Computing Study : Technology Challenges in Achieving
%  Exascale Systems}.
%\newblock {\em (P. Kogge, Editor and Study Lead)}, TR-2008-13:1--278.

%\end{thebibliography}

\end{document}